\documentclass[twocolumn]{aastex63}
\usepackage{amsmath, esint}
\usepackage{amssymb}
\usepackage{threeparttable}
\usepackage{CJK}

\def\vel{\mathbf{v}}
\def\rs{r_{\rm s}}
\def\vr{v_{\rm r}}
\def\vth{v_{\theta}}
\def\vph{v_{\phi}}
\def\kappas{\kappa_{s}}
\def\kappaa{\kappa_{a}}

\begin{document}
\begin{CJK*}{UTF8}{gbsn}

\title{A Bright First Day for Tidal Disruption Events}
\author[0000-0003-2868-489X]{Xiaoshan Huang (黄小珊)}
\affiliation{Department of Astronomy, University of Virginia, Charlottesville, VA 22904, USA}
\author[0000-0001-7488-4468]{Shane W. Davis}
\affiliation{Department of Astronomy, University of Virginia, Charlottesville, VA 22904, USA}
\author[0000-0002-2624-3399]{Yan-fei Jiang (姜燕飞)}
\affiliation{Center for Computational Astrophysics, Flatiron Institute, 162 Fifth Avenue, New York, NY 10010, USA}

\begin{abstract}
Stream-stream collision may be an important pre-peak energy dissipation mechanism in tidal disruption events (TDEs). We perform local three-dimensional radiation hydrodynamic simulations in a wedge geometry including the gravity to study stream self-crossing, with emphasis on resolving the collision and following the subsequent outflow. We find that the collision can contribute to pre-peak optical emissions by converting $\gtrsim5\%$ of stream kinetic energy to radiation, yielding prompt emission of $\sim10^{42-44}\rm erg~s^{-1}$. The radiative efficiency is sensitive to stream mass fallback rates, and strongly depends on the downstream gas optical depth. Even for a sub-Eddington ($10\%$) mass fallback rate, the strong radiation pressure produced in the collision can form a local super-Eddington region near the collision site, where a fast, aspherical outflow is launched. Higher mass fallback rate usually leads to more optically-thick outflow and lower net radiative efficiency. For $\dot{M}\gtrsim0.1\dot{M}_{\rm Edd}$, the estimated photosphere size of the outflow can expand by one to two orders of magnitudes reaching $\sim10^{14}\rm cm$. The average gas temperature at this photospheric surface is a few $\times10^{4}$K, roughly consistent with inferred pre-peak photosphere properties for some optical TDEs. We find that the dynamics is sensitive to collision angle and collision radius,  but the radiative efficiency or outflow properties show more complex dependency than is often assumed in ballistic models.
\end{abstract}

\section{Introduction}\label{sec:introduction}
The tidal disruption of a star by a black hole can power a bright transient flare in multiple wavebands \citep{rees1988tidal, phinney1989manifestations, evans1989tidal, ulmer1999flares}. These tidal disruption events (TDEs) happen when the pericenter of star's orbit is within the tidal radius $R_{\rm T} = R_{\ast}(M_{\rm BH}/M_{\ast})^{1/3}$, where $M_{\ast}$ and $R_{\ast}$ are the mass and radius of the star, and $M_{\rm BH}$ is the mass of the black hole. After disruption, roughly half of the material escapes the black hole's gravity, while other gravitationally bound gas returns to the pericenter in a highly eccentric orbit. 

The fallback and circularization of the debris can produce prominent emission, resulting in luminous transient events peaked in optical-ultraviolet (UV) or X-ray. The decay timescale of TDE candidates often approximately follows the $t^{-5/3}$ scaling, which is a theoretical prediction for the mass fallback rate of the debris \citep{rees1988tidal, phinney1989manifestations}. TDEs are important probes for studying the black hole properties such as mass \citep{ramirez2009star, stone2016rates, wevers2017black, mockler2019weighing} and spin \citep{kesden2012black, gafton2019tidal}. The chemical composition and light curve also constrain the structure and evolution stage of the disrupted star \citep{lodato2009stellar, macleod2012tidal, kochanek2016abundance, ryu2020measuring, law2020stellar}. 

TDEs were first detected as strong X-ray flares from quiescent galaxies \citep{grupe1999rx, greiner2000rx} by \textit{ROSAT}. More X-ray events have subsequently been captured by or identified in \textit{XXM-Newton}, \textit{Chandra} and \textit{Swift} \citep{lin2011discovery, saxton2012tidal,komossa2015tidal,gehrels2015journal,lin2018multiwavelength, wevers2019evidence}. Complimenting X-ray observations, the optical/UV is another important channel to detect TDEs. In the recent decade, optical transient surveys have played an increasing role in identifying TDEs, including for example the Sloan Digital Sky Survey (SDSS), Pan-STARRS and ASAS-SN surveys\citep{van2011optical, hung2017revisiting, kochanek2017all, gezari2021tidal, van2021seventeen, goldtooth2023census}. This sample is expected to greatly increase with the upcoming Large Synoptic Survey Telescope (LSST) of the Vera Rubin Observatory \citep{bricman2020prospects, gezari2021tidal, hambleton2022rubin} and Ultraviolet Transient Astronomy Satellite (ULTRASAT) \citep{ben2022scientific}.

Interestingly, the optical/UV and the X-ray identified candidates may represent different populations of emission sources. First, optical/UV bright TDEs rarely show X-ray emission. Second, the continuum temperature derived in optical/UV TDEs is usually an order of magnitude lower than the X-ray TDEs. The inferred blackbody temperature of X-ray TDEs is roughly consistent with emission from a hot accretion disk formed from the captured debris. This suggests that the optical candidates, with much lower temperature, may originate from other emission mechanisms. 

One theoretical resolution of this dichotomy involves reprocessing of X-ray emission by a quasi-spherical, optically thick layer around the accretion disk \citep{loeb1997optical, strubbe2009optical, metzger2016bright, roth2016x}, or a quasi-spherical, radiation supported disk such as the “Zero-Bernoulli accretion flows”, or ``ZEBRAs'' model \citep{coughlin2014hyperaccretion,eyles2022simulated}). In parallel, the reprocessing of disk emission could be due to an optically thick wind from the disk \citep{lodato2011multiband, miller2015disk, metzger2016bright, kara2018ultrafast}. \citet{dai2018unified} proposed a unified model for a typical TDE system, including the highly asymmetric geometry of radiation supported thick disk, optically thick wind and hot corona or jet. This model suggests that the emission from such systems will naturally cover both the optical-UV and soft X-ray bandwidth, with viewing angle effects dominating the appearance in different wavebands (also see \citet{thomsen2022dynamical}). 

The models explaining optical emission from TDEs generally involve gas circularization and radiation generation by viscous processes, which will likely to happen near or after the peak optical light curve. But observations indicate that the inferred photosphere of some TDEs experience quick expansion during the pre-peak time, where the size can increase for one to two orders of magnitudes within days to weeks \citep{mockler2019weighing, van2021seventeen}. It is likely that additional emission or reprocessing mechanism is needed to explain the pre-peak light curve for optical TDEs. 

A potentially important pre-peak emission mechanism is the stream-stream collision, which happens when the apsidal precession shifts the stream orbit and leads to stream self-intersection \citep{shiokawa2015general, piran2015disk, dai2015soft, dittmann2022analytical}. When the fallback stream and the returning stream collide, some of the streams kinetic energy will be dissipated. The intersection region can become radiation pressure dominated, if the post-shock gas is sufficiently optically thick \citep{kim1999stream,jiang2016prompt}. Consequently, the stream-stream collision can affect the pre-peak emission through different channels. First, it might directly contribute to optical emission by converting kinetic energy to radiation. Alternatively, it might create a radiation-pressure driven outflow, which reprocesses the emission from the hot, shocked gas into optical emission \citep{jiang2016prompt, bonnerot2021formation}. 

Recent global simulations that track the post-shock gas circularization seem to indicate that the stream-stream collision can efficiently redistribute debris angular momentum \citep{bonnerot2021formation, curd2021global, steinberg2022origins, andalman2022tidal}. In fact, recent work has suggested the fallback debris may be slow to circularize and form an accretion disk, implying that the majority of emission could be powered by various shocks including those from stream self interaction. The exact outcome of collision such as relative importance of luminosity and kinetic energy is sensitive to TDE system physical parameters as well as numerical treatment. \citet{jiang2016prompt} and \citet{lu2020self} performed local simulations of stream-stream collision. They show that the post-shock gas is likely to form optically-thick outflow, and the stream-stream collision itself is also an important pre-peak emission source. In this paper, we adopt similar approach and study the stream-stream collision in a localized calculation domain. The domain is optimized for resolving the streams and radiation mediated collision, while radially extended to track the post-shock gas evolution. We focus on the radiation's role in redistributing gas momentum and energy, determining the pre-peak emission, and affecting the downstream gas evolution. 

We introduce our numerical set-up in Section~\ref{sec:method_set-up}. In Section~\ref{sec:result}, we describe a case study of Eddington mass fallback rate. We also compare the effect of different mass fallback rates of the streams, the collision angle between the streams, and the collision radius relative to the black hole. Table~\ref{tab:sim_params} lists the key orbital parameters. In Section~\ref{sec:discussion}, we discuss the implied prompt emission luminosity, the estimated photosphere evolution and compare our results with previous studies. We elaborate the details of gravity implementation in Appendix~\ref{appendix:grav}, and show convergence study in Appendix~\ref{appendix:resolution}.

\section{Simulation Set-up}\label{sec:method_set-up}
\subsection{Equations and Units}
We solve the following equations in Athena++ with the explicit radiation transfer module \citep{jiang2021implicit}.

Equation~\ref{eq:gasdensity} - Equation~\ref{eq:gasenergy} are hydrodynamic equations for gas, where $\rho$, $\vel$, $E=E_{g}+(1/2)\rho v^{2}$ are fluid density, velocity and total energy density. We assume gas is ideal, so gas internal energy $E_{g}$ and pressure $P$ are related by $E_{g}=P/(\gamma-1)$. $\textsf{P}$ is the pressure tensor. In the right-hand side of Equation~\ref{eq:gasmom} and Equation~\ref{eq:gasenergy},  radiation couples with gas through the components of the radiation four force $-\mathbf{G}$ and $-cG^{0}$, which are the momentum and energy imparting from radiation to gas. We adopt the generalized Newtonian gravitational description from \citet{tejeda2013accurate}. The gravitational force and gravitational potential are  $\rho\mathbf{a}_{\rm grav}$ and $\rho\mathbf{a}_{\rm grav}\cdot\vel$, we list the detailed implementation in Appendix~\ref{appendix:grav}.
\begin{eqnarray}\label{eq:RHD}
&\frac{\partial\rho}{\partial t}+\nabla\cdot(\rho\vel)=0,\label{eq:gasdensity}\\
&\frac{\partial(\rho\vel)}{\partial t}+\nabla\cdot(\rho\vel\vel-\textsf{P})=-\mathbf{G}+\rho\mathbf{a}_{\rm grav}\label{eq:gasmom}\\
&\frac{\partial E}{\partial t}+\nabla\cdot[(E+P)\vel]=-cG^{0}+\rho\mathbf{a}_{\rm grav}\cdot\vel,\label{eq:gasenergy}\\
&\frac{\partial I}{\partial t} + c\textbf{n}\cdot\nabla I=cS_{I}\label{eq:RT}\\
&S_{I}\equiv\Gamma^{-3}[\rho(\kappas+\kappaa)(J_{0}-I_{0})\nonumber\\
&+\rho(\kappaa+\kappa_{\delta P})\left(\frac{a_{R}T^{4}}{4\pi}-J_{0}\right)]\label{eq:intensitysource}\\
&cG^{0}\equiv 4\pi c\int S_{I}d\Omega\label{eq:radenergy}\\
&\mathbf{G}\equiv4\pi\int\mathbf{n}S_{I}d\Omega\label{eq:radpressure}
\end{eqnarray}

Equation~\ref{eq:RT} - Equation~\ref{eq:radpressure} represent the frequency integrated radiation transfer equation and the source terms couple to the gas momentum and energy. In the Athena++ implementation, the specific intensities $I$ are first transformed to the fluid comoving frame, where the opacities and emissivities are the simplest. The radiation source terms $S_{I}$ is evaluated and updated implicitly along with the comoving frame gas internal energy equation. The intensities are then transformed back to the lab frame, and the resulting energy and momentum source terms $\mathbf{G}$ and $cG^{0}$ are evaluated by integrating the intensities over angle as described in \citet{jiang2021implicit}. Here, $c$ is the light speed. $I_{0}$ and $J_{0}$ are the comoving frame intensity and mean intensity. $\textbf{n}$ is the unit vector corresponding to rays in discretized angular grids, with the same angular discretization in \citet{davis2012radiation}. In the simulations, we used $n_{\mu}=3$ angles per octant, resulting $n_{\rm oct}n_{\mu}(n_{\mu}+1)/2=48$ angles in $n_{\rm oct}=8$ octants.  $\Gamma=\Gamma(\textbf{n}, \vel)=\gamma(1-\textbf{n}\cdot\vel/c)$ is the scaled Lorentz factor, with $\gamma=1/\sqrt{1-(v/c)^{2}}$ is the Lorentz factor. Equation~\ref{eq:radenergy} and \ref{eq:radpressure} represent radiation's effect on gas energy and momentum.  In the source term, $a_{R}$ is the radiation constant, $\kappas$ and $\kappaa$ are scattering opacity and Rosseland mean absorption opacity, $\kappa_{\delta P}$ is the difference between the Planck mean and Rosseland mean opacity. 

\begin{figure}
    \centering
    \includegraphics[width=\linewidth]{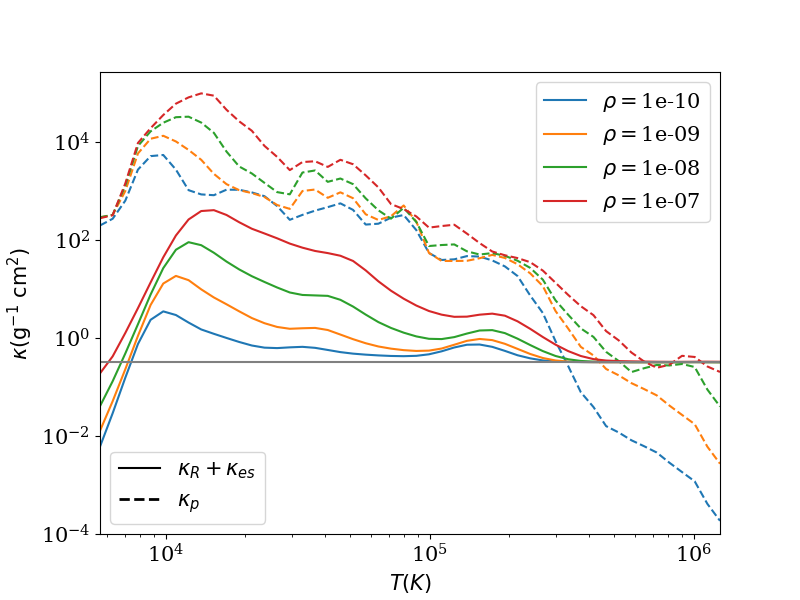}
    \caption{Examples of adopted OPAL opacity including the Planck mean opacity (dashed lines) and combined electron scattering and Rosseland mean opacity (solid lines), corresponding to densities from $\rho=10^{-10}, 10^{-9}, 10^{-8}, 10^{-7}\rm~g~cm^{-3}$ (blue, orange, green and red lines).}
    \label{fig:method_opacity}
\end{figure}
We use OPAL opacity \citep{iglesias1996updated} for the $\kappa_{\delta P}$ and $\kappaa$ based on the local gas density and temperature, and assume $\kappas=0.34~\rm cm^{2}~g^{-1}$. Figure~\ref{fig:method_opacity} shows example opacity as functions of temperature for a range of gas densities. In the code, we solve the unit-less equations with the scaling of density $\rho_{0}=10^{-10}\rm g~cm^{-3}$, velocity $v_{0}=0.01c$ and length $l_{0}=r_{\rm s}=2.95\times10^{12}$cm. In the rest of paper, we report unit-less quantities unless explicitly specified.

\subsection{Calculation Domain and Resolution}\label{subsec:method_domain}
\begin{figure}
    \centering
    \includegraphics[width=\linewidth]{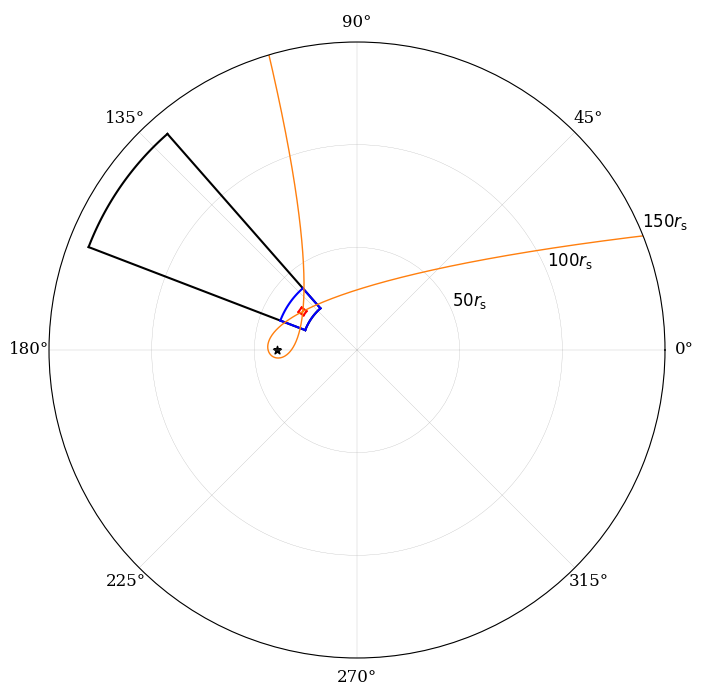}
    \caption{Schematic plot of calculation domain for a ballistic orbit (solid orange line) with $M_{\rm BH}=10^{7}M_{\odot},~M_{*}=M_{\odot},~R_{*}=R_{\odot},~\beta=1.0$. The black solid frame shows the calculation domain we adopt in A122Edd1, A122Edd01, A122Edd001 and A122Edd10. The blue solid frames shows the static mesh refinement(SMR) region with level=5. The red solid frame shows the SMR region with level=6 near the shock. The black star symbol shows the position of the black hole ($-L_{12},~\pi/2,~\pi$).}
    \label{fig:method_domain}
\end{figure}

We preform the local simulations in spherical-polar coordinates, but with the origin displaced relative to the black hole. We adopt a simulation domain that centers the collision point in the $\phi$ direction, and varies with different stream orbits.  As an example, Figure~\ref{fig:method_domain} shows the domain we adopt for A122 runs (A122Edd1, A122Edd01, A122Edd001 and A122Edd10).  As we will elaborate in Section~\ref{subsec:result_resolution}, resolving the streams and the collision region is the primary motivation for us to adopt such a local simulation domain. In the rest of the section, we note the variables in the coordinate with the black hole at the origin with subscript 1. We first integrate $\ddot{r_1},~\ddot{\theta_1}$ and $\ddot{\phi_1}$ from \citet{tejeda2013accurate} in the $\theta=\pi/2$-plane to obtain the ballistic trajectory with stream-stream collision radius $r_{\rm coll,1}$ and collision angle $\phi_{\rm coll,1}$ with respect to the black hole. We set the black hole and star mass to $M_{\rm BH}=10^{7}M_{\odot}$ and $M_{*}=M_{\odot}$ respectively, and the stellar radius is $R_{*}=R_{\odot}$. We assume the orbit eccentricity $e_{\rm orb}$ estimated as \citet{dai2015soft}
\begin{equation}\label{eq:ecc_orb}
    e_{\rm orb}\approx 1-0.02\left(\frac{M_{*}}{M_{\odot}}\right)^{1/3}\left(\frac{M_{\rm BH}}{10^{6}M_{\odot}}\right)^{-1/3}\beta^{-1},
\end{equation}
where $\beta=r_{\rm T}/r_{\rm p}$ is the orbit penetration factor. $r_{\rm T}=R_{*}(M_{\rm BH}/M_{*})^{1/3}$ is the tidal radius, and $r_{\rm p}$ is the pericenter radius. So by choosing $\beta$, we obtain the initial condition for a Newtonian elliptical orbit as an approximation to the orbit of most bound material. The initial position and velocity are $(r_{\rm SMA},~\pi/2.0,~0.0)$ and  $(0.0,~0.0,~v_{\phi,\rm SMA})$ in $r,~\theta,~\phi$ direction, here $r_{\rm SMA}=r_{\rm p}/(1-e_{\rm orb})$ is the approximated semi-major axis, $v_{\phi,\rm SMA}=\sqrt{(GM_{\rm BH}/r_{\rm SMA})(1-e_{\rm orb})/(1+e_{\rm orb})}$ is the approximated initial $\phi$ direction velocity. The solid orange line in Figure~\ref{fig:method_domain} shows part of the orbit with $\beta=1.0$, $r_{\rm SMA}\approx1088\rs$, $v_{\phi, \rm SMA}\approx0.002c$. With this orbit, the streams collide at $r_{\rm coll,1}=22.5\rs$ relative to the black hole with the colliding angle $\theta_{\rm coll,1}=122^{\circ}$. 

The domain is extended in $r$ direction, to align with the net momentum direction of the collision. Instead of placing the black hole at the origin, we translate the black hole by $L_{12}$, so it is at ($-L_{12},~\pi/2,~\pi$). For A122 runs, $L_{12}=r_{\rm coll,1}/\cos\phi_{\rm coll,1}=39.15\rs$. For A90Edd1 and A122R95Edd1, we chose $L_{12}=69.48\rs$ and $L_{12}=118.25\rs$. In the $r$ direction, we use logarithmic grid spacing. For A122 runs, the domain is $(27\rs,140\rs)\times(0.42\pi,0.58\pi)\times(0.81\pi,0.96\pi)$ in $r,~\theta,~\phi$ direction, which is shown in Figure~\ref{fig:method_domain} by the black frame as the projection on $\theta=\pi/2$-plane. For A90Edd1, the domain is $(40\rs,207\rs)\times(0.42\pi,0.58\pi)\times(0.81\pi,0.97\pi)$. For A122R95Edd1, the domain is $(44\rs,228\rs)\times(0.42\pi,0.58\pi)\times(0.64\pi,0.79\pi)$.

We use static mesh refinement (SMR) in Athena++ to resolve the streams and the shock. The root level is resolved by $[64\times32\times64]$ cells in $r,~\theta,~\phi$ direction directions. The blue and red regions in Figure~\ref{fig:method_domain} show the refined regions in A122Edd1, A122Edd01, A122Edd001 and A122Edd10. The blue region is at five levels of refinement and spans $(27\rs,40\rs)\times(0.48\pi,0.52\pi)\times(0.81\pi,0.96\pi)$, it covers most of the stream trajectory before the stream-stream collision, giving the resolution $\delta R\approx R\delta\theta\approx0.02\approx0.85R_{\odot}$ in $r$ and $\theta$ directions. The red region is at six levels of refinement and is $(31\rs,34\rs)\times(0.49\pi,0.51\pi)\times(0.82\pi,0.97\pi)$. For A90Edd1, the level five refinement region is $(44\rs,60.5\rs)\times(0.42\pi,0.58\pi)\times(0.81\pi,0.97\pi)$, and level six refinement region is $(56.5\rs,59.5\rs)\times(0.42\pi,0.58\pi)\times(0.88\pi,0.91\pi)$. For A122R95Edd1, the level five refinement region is $(45.0\rs, 58.0\rs)\times(0.42\pi,0.58\pi)\times(0.64\pi,0.79\pi)$, and level six refinement region is $(49.5\rs,54.0\rs)\times(0.42\pi,0.58\pi)\times(0.70\pi,0.73\pi)$. 

At the collision radius $r_{\rm coll,1}$, assuming the stream scale height $H$ is determined by the vertical balance between tidal force and gas pressure gradient \citep{guillochon2014ps1, jiang2016prompt}, we estimate that $H\approx2r_{\rm coll,1}\sqrt{(r_{\rm coll,1}/\rs)(k_{\rm B}T_{\rm str}/\mu c^{2})}$.  With a $M_{\rm BH}=10^{7}M_{\odot}$ black hole and $r_{\rm coll,1}\approx20\rs$ orbit, $H\approx3R_{\odot}\sqrt{T_{\rm str}/10^{6}K}$. With five levels of SMR, we can resolve the $H$ by roughly four cells, assuming stream temperature $T_{\rm str}=10^{6}$K before collision. Similar to \cite{jiang2016prompt}, we assume higher $T_{\rm str}$ to calculate the scale height, but set the gas temperature in the stream to be $T_{\rm str}\approx4\times10^{5}$K. Before collision, the kinetic energy density in the streams is about six orders of magnitude higher than internal energy density or radiation energy density. After the collision, the post-shock gas temperature is mostly in equilibrium with effective radiation temperature. As we will discuss in Section~\ref{sec:result}, the post-shock gas radiation energy density is primarily determined by the energy conversion from kinetic energy. Therefore, the internal energy in the stream prior to collision has limited effect on post-shock gas temperature.

\subsection{Initial and Boundary Condition}\label{subsec:method_initbval}
Before injecting stream, we set the initial background density and pressure to be low values of $\rho_{\rm init}=2.7\times10^{-5}$ and $P_{\rm init}=2.65\times10^{-9}$. We set density and pressure floor for the hydrodynamic Riemann solver to be $\rho_{\rm floor}=2\times10^{-5}$ and $P_{\rm floor}=2\times10^{-9}$. 

We model the stream-stream collision as a fallback stream and a returning stream injected from both $\phi$ boundaries in the $\theta=\pi/2$ plane, similar to \citet{jiang2016prompt}. For each stream, we inject them at the radius of $r_{\rm inj}$ and $\theta_{\rm inj}=\pi/2$ by marking the four nearest ghost cells in $r$ and $\theta$ direction as stream injection cells, and set uniform gas density $\rho_{\rm inj}$, velocity $\textbf{v}_{\rm inj}$ and pressure in these cells. We do not explicitly set the radiation intensity in ghost cells, once the gas streams enter calculation domain, they rapidly adjust to thermal equilibrium in just a few zones. The other cells at the $\phi$ direction boundaries are set to be single direction outflow for hydrodynamical variables, which copies all the values from the first active cells but set any velocity that enters the calculation domain to zero. The $r$ and $\theta$ direction hydro boundaries are also single direction outflow. The radiation boundaries in the $r$, $\theta$ and $\phi$ direction are ``vacuum'' radiation boundaries, which copies all the intensities with $\textbf{n}$ pointing outward, but sets all intensities with $\textbf{n}$ pointing inward to be zero.

For the stream injection cells at the $\phi$ direction boundaries, we set local velocity to be $\textbf{v}_{\rm inj}=(v_{\rm r,inj}, ~0, ~v_{\rm \phi, inj})$. The density in the stream injection ghost cells is set by $\rho_{\rm inj}=f_{\rm Edd}\dot{M}_{\rm Edd}/(v_{\rm \phi, inj}A_{\phi,\rm inj})$, where $\dot{M}_{\rm Edd}=40\pi GM_{\rm BH}/(C\kappas)$ is the Eddington accretion rate, $f_{\rm Edd}$ is the Eddington ratio, and $A_{\phi,\rm inj}\approx H^{2}$ is the total injection area normal to $\phi$ direction. In simulations, $A_{\phi,\rm inj}$ varies as we change $r_{\rm inj}$ and $\phi_{\rm inj}$, so we use the numerically computed cell areas for better precision.

Note that in our set-up, we assume the fallback stream and the returning stream have the same entropy and ignore the potential stream width change due to the dissipation near the pericenter. We discuss the potential effects in Section~\ref{subsec:discussion_compare}, but leave the study of how nozzle shock dissipation, radiative cooling and numerical diffusion affect stream width after pericenter to future work. 

To obtain $r_{\rm inj}$ and $\textbf{v}_{\rm inj}$, we first integrate Equation~\ref{eq:rosswog} to get the ballistic trajectory and velocity of an assumed point mass, then translate them by $L_{12}$. For A122Edd1, A122Edd01, A122Edd001 and A122Edd10, $r_{\rm inj}=29.4\rs,~29.7\rs$, $\textbf{v}_{\rm inj}=(0.0534c,~0,~0.167c),~(0.0343c,~0,~0.235c)$ at the inner and outer $\phi$ direction boundary. For A122Edd1, $\rho_{\rm inj}=7.076\times10^{-8}\rm g~cm^{-3},~4.974\times10^{-8}\rm g~cm^{-3}$ at the inner and outer $\phi$ direction boundary to match $\dot{M}_{\rm Edd}$. The $\rho_{\rm inj}$ of A122Edd01, A122Edd001 and A122Edd10 are $1\%, 10\%, 1000\%$ of A122Edd1.

For A90Edd1, we keep the fallback stream the same as A122Edd1, but manually change the returning stream velocity so that it collides with the fallback stream at $r_{\rm coll,1}=24.1\rs$ and $\theta_{\rm coll,1}=90^{\circ}$. We inject the streams at $r_{\rm inj}=47.74\rs,~52.89\rs$, with the density and velocity $\textbf{v}_{\rm inj}=(0.092c,~0.0,~0.12c),~(0.0065c,~0.0,~0.18c)$ at the inner and outer $\phi$ direction boundary. Similarly, for A122R95Edd1, the fallback stream is a part of the same orbit as in A122Edd1, but we manually change the returning stream, so they collide with similar angle $\theta_{\rm coll,1}=125^{\circ}$ but at a further radius $r_{\rm coll,1}=95.1\rs$. We inject the streams at $r_{\rm inj}=47.94\rs,~46.87\rs$, with the density and velocity $\textbf{v}_{\rm inj}=(2.24\times10^{-2}c,~0.0,~8.81\times10^{-2}c),~(3.18\times10^{-2}c,~0.0,~0.115c)$ at the inner and outer $\phi$ direction boundary respectively.

\section{Results}\label{sec:result}
We present six simulations, including four A122 runs with the same orbit but different fallback rates, along with two runs to study the effect of the collision angle (A90Edd1) and radius (A122R95Edd1). We list the stream orbital parameters in the black hole-centered coordinate Table~\ref{tab:sim_params}.
\begin{table}
\caption{Summary of Parameters for Main Simulations}
\label{tab:sim_params}
\begin{threeparttable}
\centering
\begin{tabular}{lccc}
\hline
Name & $\dot{M}/\dot{M}_{\rm Edd}$ & $\theta_{\rm coll,1}$ & $r_{\rm coll,1}$\\
\hline
A122Edd1 & 1.0 & $122^{\circ}$ & 22.5$\rs$\\
\hline
A122Edd01 & $10^{-1}$ & $122^{\circ}$ & 22.5$\rs$\\
\hline
A122Edd001 & $10^{-2}$ & $122^{\circ}$ & 22.5$\rs$\\
\hline
A122Edd10 & $10.0$ & $122^{\circ}$ & 22.5$\rs$\\
\hline
A90Edd1 & 1.0 & $90^{\circ}$ & 24.1$\rs$\\
\hline
A122R95Edd1 & 1.0 & $125^{\circ}$ & 95.1$\rs$\\
\hline
\end{tabular}
\end{threeparttable}
\end{table}

\subsection{Importance of resolving the stream and radiation mediated shock}\label{subsec:result_resolution}
In this section, we show that the outcome of stream-stream collision in RHD simulations is highly sensitive to spatial resolution. We compare a low resolution simulation A122Edd1\_LR and the fiducial resolution A122Edd1, where we reduce the level-five and level-six refinement to level-two refinement. With the adjustment, we have the resolution of $\delta R\approx R\delta\theta\approx0.16$ in $r$ and $\theta$ direction when injecting the streams, meaning $\delta R$ is about $1/8$ times of the fiducial A122Edd1 resolution at the same location. So we no longer resolve the estimated stream scale height $H$. As a result, the stream cross-section $A\sim\delta R^{2}$ is artificially increased, so the stream density $\rho_{\rm inj}\sim \dot{M}/Av_{\rm inj}\propto\delta R^{-2}$ is reduced at fixed $\dot{M}$. We set $\rho_{\rm inj}=1.10\times10^{-9}\rm g~cm^{-3},~7.73\times10^{-10}\rm g~cm^{-3}$ at the inner and outer $\phi$ direction boundary to match $f_{\rm Edd}=1.0$ fallback rate. Other simulation set-ups are the same as A122Edd1.

The choice of level-two refinement in A122Edd1\_LR  is motivated by the global test calculation in Appendix~\ref{appendix:grav}, which tracks the stream orbit around the black hole. In the test simulation, we observe noticeably different gas dynamics after the stream-stream collision compared to A122Edd1, despite the same stream orbit and comparable mass injection rates.

In A122Edd1\_LR, the stream density is artificially reduced due to insufficient resolution, which could affect the stream vertical optical depth, where $\tau_{\rm str}\sim\kappa_{\rm s}\rho_{\rm inj}\delta R\propto\delta R^{-1}$. We measure the initial vertical optical depth $\tau_{\rm str}$ by integrating $\tau_{\rm str}=\int \kappas\rho dr$ from $r=29.0$ to $r=31.0$ at $\phi=0.86\pi$ in $r$ direction. We find that $\tau_{\rm str}\approx6\times10^{2}$ in A122Edd1\_LR, and $\tau_{\rm str}\approx5\times10^{3}$ in A122Edd1. So before the collision, both streams are optically thick, but with order-of-magnitude different vertical optical depth. 

\begin{figure}
    \centering
    \includegraphics[width=\linewidth]{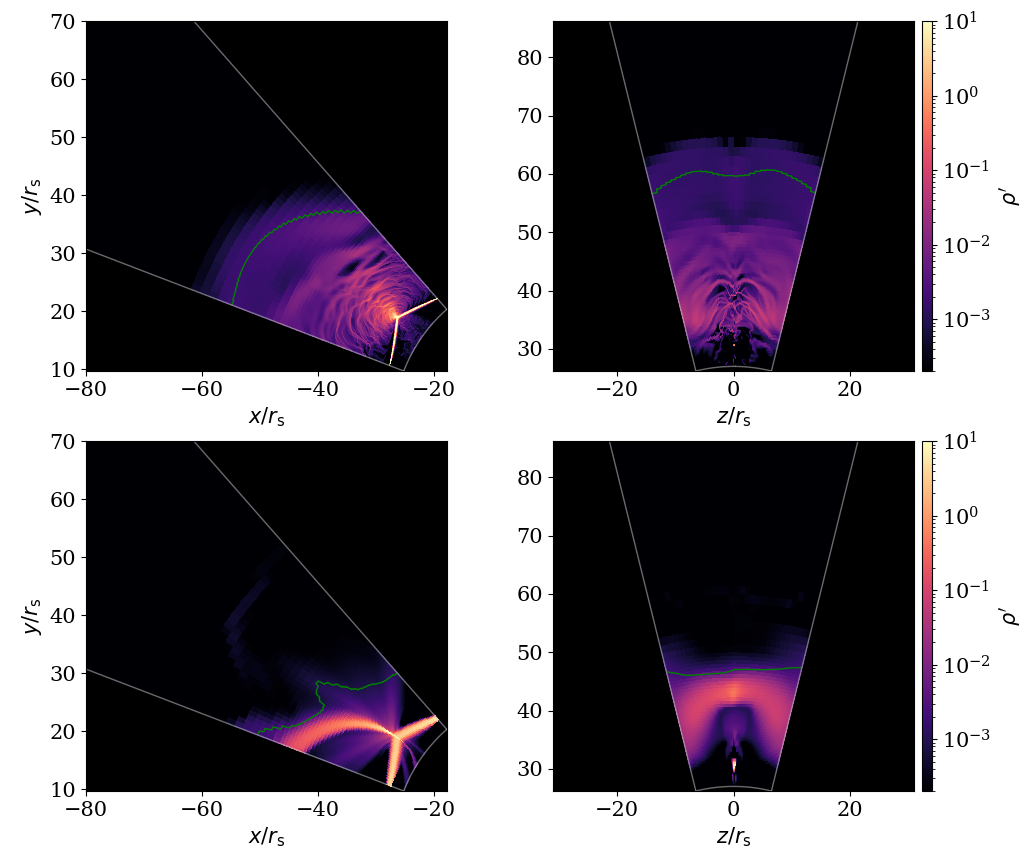}
    \caption{Gas density snapshots at $t=0.3$ for A122Edd1 (the upper panel) and A122Edd1\_LR (the lower panel), where A122Edd1 is the fiducial resolution simulation with $\delta R\approx0.02$ (in unit of $\rs$), and A122Edd1\_LR is the low resolution simulation with $\delta R\approx0.16$ (in unit of $\rs$) when the streams are injected. In each row, the left plot shows volume average from $\phi=1.56-1.59$ near $\theta=\pi/2$-plane, the right plot shows volume average from $\theta=2.52-2.55$ near mid-$\phi$-plane.}
    \label{fig:result_resolution_density}
\end{figure}

Figure~\ref{fig:result_resolution_density} shows the distinct downstream gas distribution in A122Edd1\_LR and A122Edd1. After the two streams collide, the gas in A122Edd1 forms an expanding outflow that propagates radially outward in the orbital plane. Off the orbital plane, gas moves away symmetrically with respect to $\theta=\pi/2$-plane. The expanded downstream gas distribution suggests that the radiation pressure produced in the collision redistributes the gas. In contrast, post-shock gas in A122Edd1\_LR is mainly concentrated in a merged stream that bends towards the black hole. 

The different outflow morphology suggests different radiative acceleration near the shock in the two simulations. We find that the radiation energy density at stream-stream collision shock front in A122Edd1\_LR is about two orders of magnitude lower than A122Edd1 (see Appendix~\ref{appendix:resolution}). This implies that less stream kinetic energy is converted to radiation energy locally by the collision, or more efficient cooling due to enhanced radiative diffusion resulting from the lower optical depth. These effects may be associated with the lower stream kinetic energy density and the less-resolved collision shock in A122Edd1\_LR. Consequently, we find that the radiation pressure near the collision is lower in A122Edd1\_LR , which is less likely to drive massive outflow. 

The downstream gas optical depth is also different. We estimate the photosphere surface radius $R_{\rm \tau=1}$ in the $r$ direction as the radius at which :
\begin{equation}\label{eq:tau1}
    \int_{R_{\rm \tau=1}}^{R_{\rm out}}\kappas\rho dr=1,
\end{equation}
where ${R_{\rm out}}$ is the outer boundary radius the $r$ direction. The green line in Figure~\ref{fig:result_resolution_density} of each panel shows $R_{\rm \tau=1}$ of these snapshots. At $t=0.3$, average $R_{\rm \tau=1}$ of A122Edd1 is generally larger than A122Edd1\_LR. Hence, radiation diffusion takes longer and there is less cooling in the expanding outflow, the gas temperature at  $R_{\rm \tau=1}$ is higher in A122Edd1\_LR.

The primary energy conversion in the simulation is between gas kinetic energy, gas gravitational potential and radiation energy. \citet{tejeda2013accurate} defines the total conserved energy
\begin{eqnarray}\label{eq:EG}
    E_{\rm G}\equiv\frac{1}{2}\left[\frac{r^{2}v_{ r}^{2}}{(r-\rs)^{2}}+\frac{r}{r-\rs}(v_{\theta}^{2}+v_{\phi}^{2})\right]-\frac{GM_{\rm BH}}{r}
\end{eqnarray}
In the rest of paper, we define  ``orbital energy'' as $E_{\rm G}$, ``kinetic energy'' as non-relativistic kinetic energy, and ``gravitational potential'' as the difference between $E_{\rm G}$ and kinetic energy. 

\begin{figure}
    \centering
    \includegraphics[width=\linewidth]{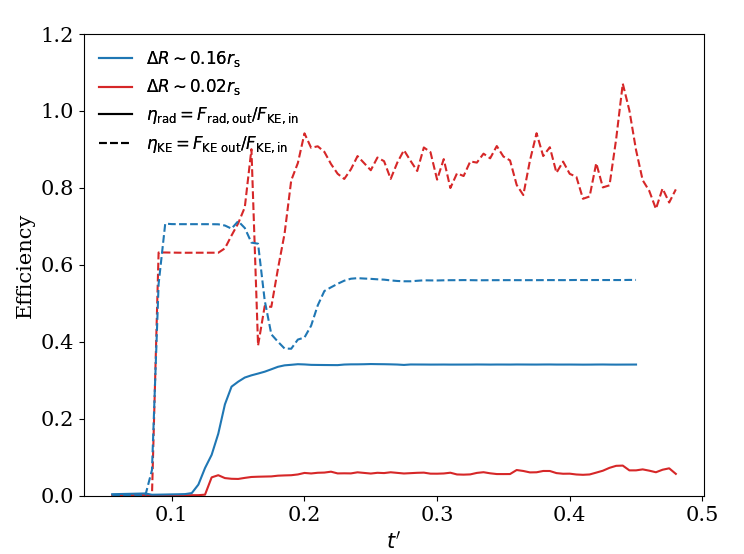}
    \caption{The radiative efficiency (the solid lines) and kinetic efficiency (the dashed lines) defined as Equation~\ref{eq:efficiency}. The blue lines are A122Edd1\_LR, and the red lines are A122Edd1. The efficiencies may slightly exceed 1.0 because of the conversion between kinetic energy and gravitational potential. The low resolution run A122Edd1\_LR produces larger radiation luminosity than A122Edd1 and less kinetic energy.}
    \label{fig:result_resolution_efficiency}
\end{figure}

In the optically-thick outflow of A122Edd1, the radiation force continuously accelerates gas until photons diffuse out from photosphere, so some radiation energy produced in the collision will be converted back to gas kinetic energy. To quantify the energy conversion, we define the net radiative efficiency $\eta_{\rm rad}$ and kinetic energy efficiency $\eta_{\rm KE}$ as:
\begin{equation}\label{eq:efficiency}
    \eta_{\rm rad}=\frac{\int\mathcal{F}_{\rm rad,out}dA}{\int\mathcal{F}_{\rm KE,in}dA},~\eta_{\rm KE}=\frac{\int\mathcal{F}_{\rm KE,out}dA}{\int\mathcal{F}_{\rm KE,in}dA},
\end{equation}
where $\mathcal{F}_{\rm KE,in}$, $\mathcal{F}_{\rm KE,out}$, $\mathcal{F}_{\rm rad,out}$ are total injected kinetic energy flux, total outgoing kinetic energy flux and total outgoing radiation flux. The fluxes are integrated on all the surfaces of an enclosed volume.

Figure~\ref{fig:result_resolution_efficiency} shows the efficiencies as a function of time for A122Edd (red lines) and A122Edd\_LR (blue lines). At $t\approx0.12$, the streams collides, the radiation energy efficiency start to rise, followed by the kinetic energy efficiency drop. Later, as radiation forces accelerate gas and downstream gas gains kinetic energy, the kinetic energy efficiency rises again, eventually relaxing to the quasi-steady-state value. Compared to A122Edd1, more injected kinetic energy is converted to radiation in A122Edd1\_LR. This is because the radiation produced in the collision diffuses through the more optically thin downstream gas without as much interaction with gas, yielding higher net radiative efficiency. Notice that the efficiencies are calculated from energy fluxes injecting and leaving the boundary surfaces of an enclosed volume within the calculation domain. Here we integrate over the surfaces of the simulation domain up to $r=40.0$, equivalent to all the surfaces of the volume $(27,40)\times(0.42\pi,0.58\pi)\times(0.81\pi,0.96\pi)$. We also calculated the efficiencies at the enclosed volume with outer radii of $r=45.0,~50.0$ and did not find a significant difference. 

When the resolution is insufficient, the reduced radiation production from the stream-stream collision may suppress the prompt luminosity, but the low optical depth in the post-shock gas may allow photons to diffuse out more easily, enhancing the prompt luminosity. Considering these competing factors, it is possible that in other sets of parameters, insufficient resolution might lead to underestimating luminosity if the effect of downstream gas optical depth is secondary. 

\subsection{Eddington mass flux: A122Edd1 }\label{subsec:restult_fiducial}
\begin{figure*}
    \centering
    \includegraphics[width=\textwidth]{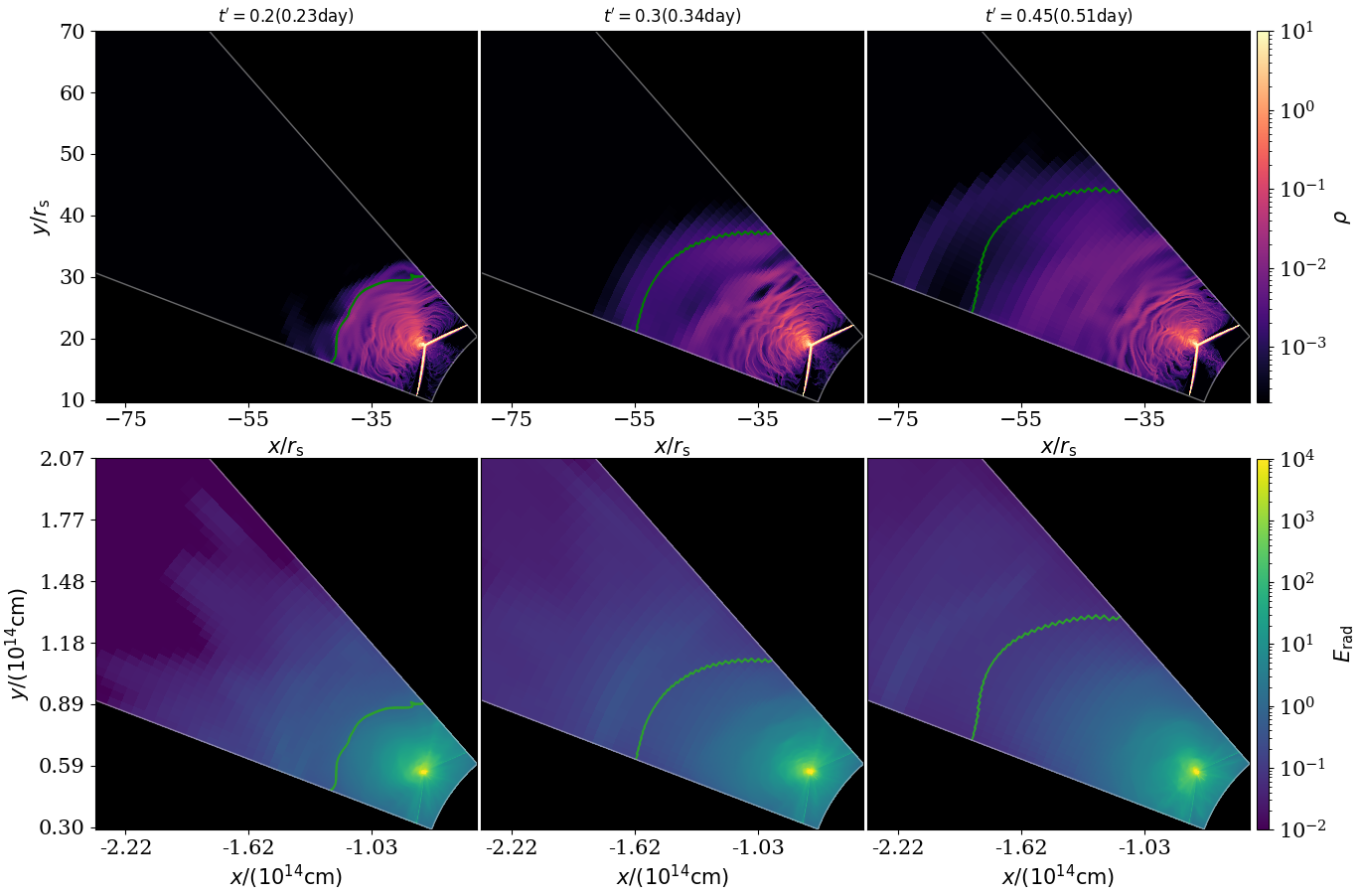}
    \caption{Gas density snapshots (the upper panel) and lab frame radiation energy density snapshots (the lower panel) at $t=0.2,~0.3,~0.45$ (from left to right) of A122Edd1. The plotted variables are volume average from $\theta=1.56-1.59$ near $\theta=\pi/2$-plane. In the lower panel, the ray effect near the collision point is likely due to angular discretization. In each plot, the green solid line is the photosphere surface as defined in Equation~\ref{eq:tau1}. }
    \label{fig:Result_edd1_dens}
\end{figure*}

\begin{figure*}
    \centering
    \includegraphics[width=\textwidth]{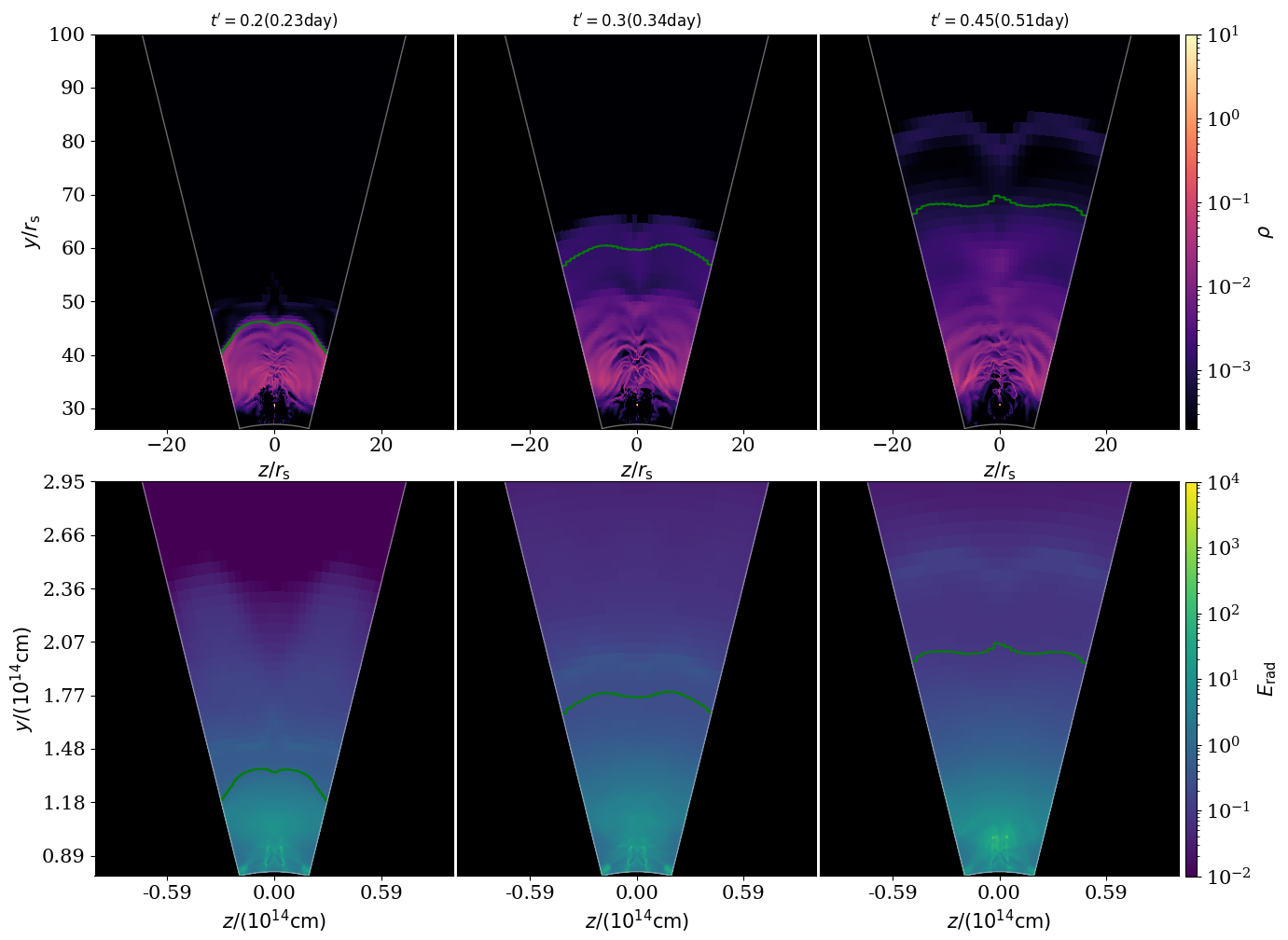}
    \caption{Gas density snapshots (the upper panel) and lab frame radiation energy density snapshots (the lower panel) at $t=0.2,~0.3,~0.45$ for A122Edd1 respectively (from left to right). The plotted variables are volume average from $\phi=2.52-2.55$ near mid-$\phi$-plane. In the lower panel, the ray effect near the collision point is likely due to angular discretization. In each plot, the green solid line is the photosphere surface as defined in Equation~\ref{eq:tau1}. }
    \label{fig:Result_edd1_dens_v}
\end{figure*}

Figure~\ref{fig:Result_edd1_dens} and Figure~\ref{fig:Result_edd1_dens_v} show the density snapshots from A122Edd1. After two streams collide, the downstream gas is accelerated by the radiation pressure, forming an optically-thick outflow that extends a few tens of $\rs$. The post-shock gas forms multiple shells near the collision site. The shell morphology is similar to \citet{jiang2016prompt}, which is related to the shock-front obliqueness changing due to slightly unbalanced pressure. 

\begin{figure}
    \centering
    \includegraphics[width=\linewidth]{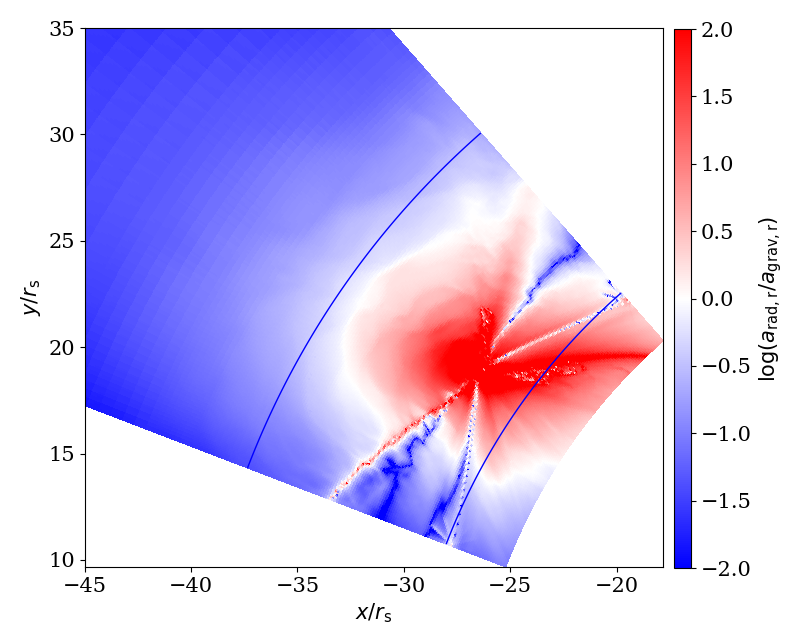}
    \caption{The ratio between radiation acceleration and gravity acceleration in $r$ direction, averaged over $\pi/2-0.24\leq\theta \pi/2+0.24$ near the $\theta=\pi/2$-plane. The two blue solid lines labels the radius $30\rs$ and $40\rs$. Due to the radiation pressure near the collision site,  the $r$ direction radiation force exceeds gravity in the downstream gas, creating a local super-Eddington region that extends few $\rs$ from the collision.}
    \label{fig:RESULT_edd1force}
\end{figure}

We find that the downstream gas is aspherical relative to the collision point and the black hole. Most gas stays within a cone of finite solid angle. For example, at $t=0.3$, gas mainly stays in $25^{\circ}\lesssim\theta_{3}\lesssim150^{\circ},~100^{\circ}\lesssim\phi_{3}\lesssim200^{\circ}$, where $\theta_{3}$, $\phi_{3}$ are the altitude and azimuthal angle relative to the collision point. In the rest of paper, we use subscript $3$ to note the variables relative to the collision point. Our calculation domain only covers part of the sky seen from the black hole, and the gas roughly fills the angular space of $60^{\circ}\lesssim\theta_{1}\lesssim120^{\circ},~55^{\circ}\lesssim\phi_{1}\lesssim140^{\circ}$ seen from the black hole. This picture is qualitatively similar to the estimated outflow morphology in \citet{lu2020self}.
The bottom panels in Figure~\ref{fig:Result_edd1_dens} and Figure~\ref{fig:Result_edd1_dens_v} show that the collision as a bright point on radiation energy density map. The radiation energy density is enhanced by about four orders of magnitudes (also see Figure~\ref{fig:appendix_streamprofile}) at the collision, where gas experiences eruptive acceleration. Figure~\ref{fig:RESULT_edd1force} shows the ratio between radiation acceleration and gravity in the $r$ direction. Even if the mass fallback rate is just Eddington, the strong radiation pressure produced in the collision creates a super-Eddington that region extends a few $\rs$ in the downstream gas. 

As gas is accelerated by radiation pressure and propagates outward, it forms an optically-thick outflow. The green solid line in each density snapshot shows the photosphere radius $R_{\tau=1}$ (Equation~\ref{eq:tau1}). The average $R_{\tau=1}\approx80\rs$ in the quasi steady state, and is relatively uniform in $\theta$ and $\phi$ direction. Along $\theta$ and $\phi$ direction line of sight, however, the simulation does not capture the equivalent photosphere surface. The gas is partially optically thick at the $\theta$ and $\phi$ boundary surfaces. We note that $R_{\tau=1}$ is defined relative to the origin, instead of the black hole. We also calculated the $F_{\rm rad}/cE_{\rm rad}$ in lab frame, where $F_{\rm rad}$ is the radiation flux magnitude, and $E_{\rm rad}$ is the radiation energy density as another metric to measure the opaqueness. We compare the surface where $F_{\rm rad}/cE_{\rm rad}=0.3, 0.4$ with $R_{\tau=1}$ and find rough agreement. Therefore, the gas outflow in A122Edd1 is optically thick and could potentially correspond to a reprocessing layer for optical TDEs at early time.

As the radiation force does work on the downstream gas, part of the gas gains energy and becomes unbound. Similar to \citet{jiang2014global}, we define unbound gas as with $E_{\rm t}>0$, where $E_{\rm t}$ is analogue to Bernoulli number:
\begin{equation}
    E_{\rm t}=E_{\rm G}+\frac{\gamma P}{\gamma-1}+\frac{4E_{\rm rad}}{3},
\end{equation}
where $E_{\rm G}$ is the total orbital energy defined in Equation~\ref{eq:EG}. The second term is fluid enthalpy, and the third term assumes radiation acts as a fluid with adiabatic index of $\gamma=4/3$. Hence, when radiation diffusion is present, this enthalpy is not a conserved quantity. The red solid line in Figure~\ref{fig:mdot_massrate} shows the total unbound mass flux, normalized to the total injected mass flux carried by the original stellar streams. As the collision happens near $t=0.18$, the unbound mass flux quickly increases to $\sim15\%$, and eventually reaches quasi-steady state, suggesting that $\sim18\%$ of injected mass flux becomes unbound when they leave the simulation domain. We found the unbound gas spatial distribution is anisotropic. Near the collision, the unbound gas has a mass-weighted average radial velocity of  $v_{\rm ub, r3}\approx 0.17c$ at quasi-steady state relative to the collision, and decreasing to  $v_{\rm ub, r3}\lesssim 0.1c$ away from the collision point outward. Therefore, the collision in A122Edd1 is able to create fast and relatively dens unbound gas. But we also note that this  ``unbound'' criteria $E_{\rm t}$ is sensitive to the stream initial condition. Unbound gas could also have orbits oriented towards the black hole. Since the initial stream is itself weakly bound, the delineation between bound and unbound gas after the collision will depend on the details of the global evolution.

\subsection{Effect of Eddington ratio}\label{subsec:result_mdot}
To study the effect of mass fallback rate, we compare the simulations A122Edd001, A122Edd01 and A122Edd10 with the fiducial simulation A122Edd1. The range of Eddington ratio provides us with a sample of different TDE systems, as well as potential different time stamps before the peak, where the mass feeding rate may rise from sub-Eddington to super-Eddington.

\begin{figure}
    \centering
    \includegraphics[width=\linewidth]{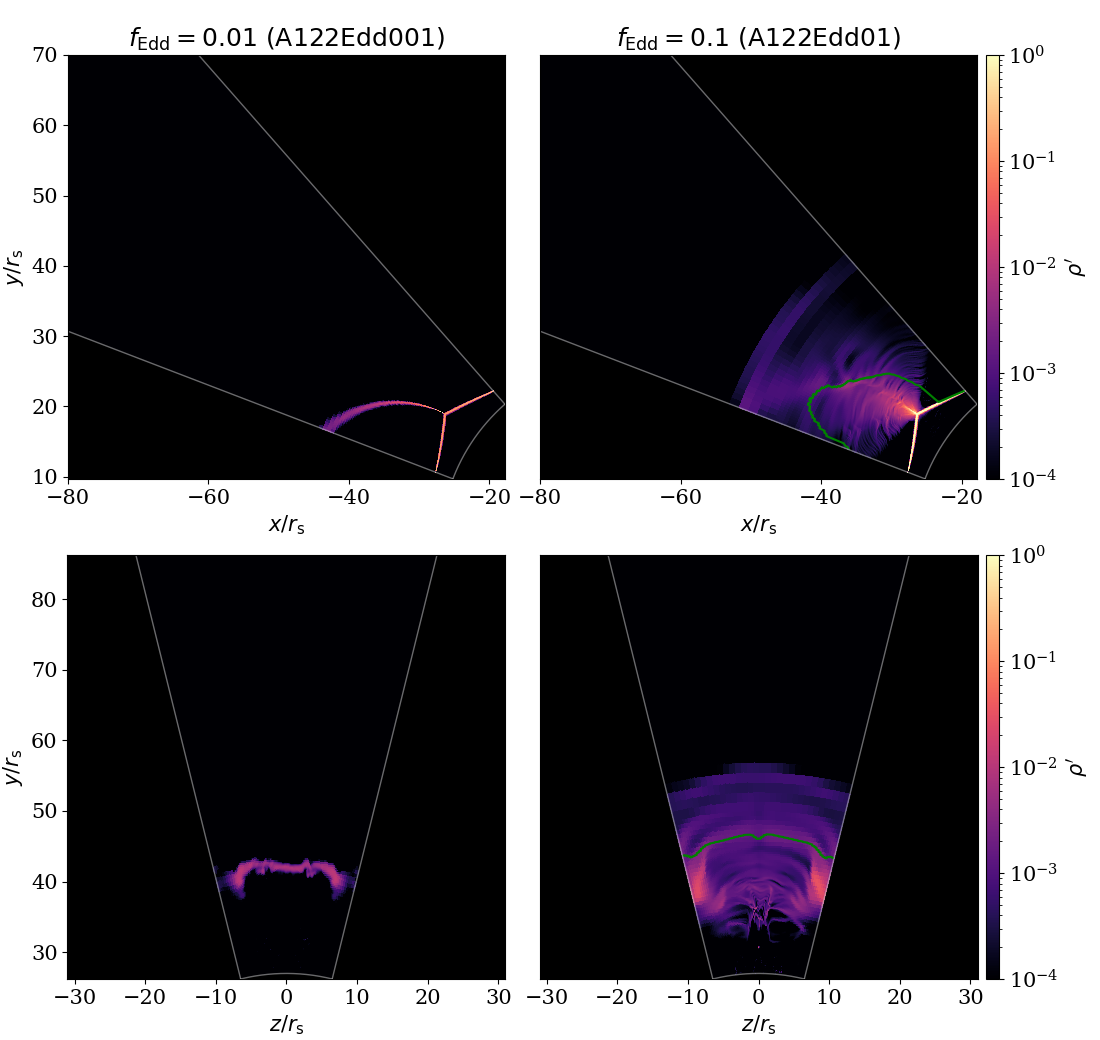}
    \caption{Gas density snapshots at $t=0.3$ for A122Edd001 (the left column) and A122Edd01 (the right column). In each column, the upper panel shows the top view (volume average from $\theta=1.56-1.59$ near $\theta=\pi/2$-plane), the lower panel shows the side view (volume average from $\phi=2.52-2.55$ near mid-$\phi$-plane). The green solid line is the photosphere surface as defined in Equation~\ref{eq:tau1}. When comparing with the fiducial simulation A122Edd1 (Figure~\ref{fig:Result_edd1_dens} and Figure~\ref{fig:Result_edd1_dens_v}), notice the different color normalization in this plot to better show low density gas.}
    \label{fig:lowedd_dens}
\end{figure}

We find that the outcome of collision is sensitive to the fallback rate, especially for sub-Eddington rates. The super-Eddington simulation A122Edd10 shows qualitatively similar outflow as A122Edd1. In contrast,  when we lower the mass fallback ratio to $10\%$ and $1\%$ Eddington, the downstream gas distribution is changed significantly. Figure~\ref{fig:lowedd_dens} shows gas density distribution from the two lower Eddington ratio simulations. In A122Edd001, the downstream gas is mainly confined in a thin stream and does not form an expanded outflow. In A122Edd01, the stream-stream collision redistributes gas to form an outflow, but this outflow has lower density when compare to the A122Edd1, and is also slightly bent towards the black hole.

In the simulations with different mass fallback rates, we assume the stream specific kinetic energy is the same. So lower mass fallback rate yields less available kinetic energy before collision. We find that compared to A122Edd1, the maximum radiation energy density at the shockfront is $0.55\%$ in A122Edd001, $9.6\%$ in A122Edd01, and $405.0\%$ in A122Edd10, measured at $t'=0.4$. Due to lower radiation pressure in sub-Eddington runs, they are less likely to drive massive outflows.

\begin{figure}
    \centering
    \includegraphics[width=\linewidth]{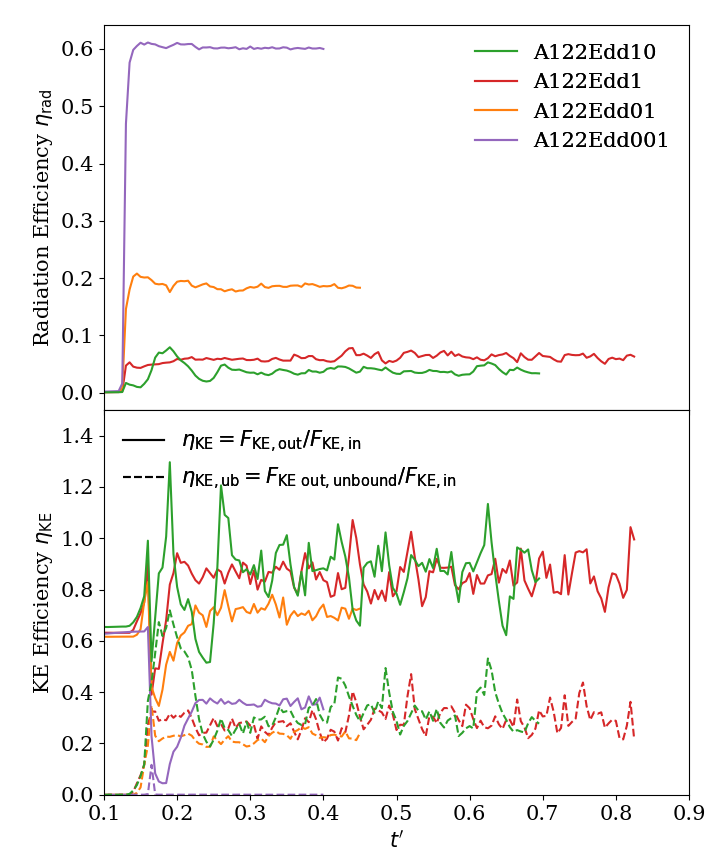}
    \caption{radiative efficiency $\eta_{\rm rad}$ (upper panel) and kinetic energy efficiency $\eta_{\rm KE}$ (lower panel, Equation~\ref{eq:efficiency}) as a function of time. In the lower panel, the solid lines are total kinetic energy leaving simulation domain, the dashed lines are unbound kinetic energy fraction. The kinetic energy efficiency might be larger than 1.0 because of conversion with gravitational potential.}
    \label{fig:efficiency_massrate}
\end{figure}

Figure~\ref{fig:efficiency_massrate} shows the radiative efficiency $\eta_{\rm rad}$ and kinetic energy efficiency $\eta_{\rm KE}$. It shows the trend that higher Eddington ratio leads to lower radiative efficiency. We attribute the anti-correlation to energy exchange between radiation and kinetic energy in downstream gas. To first order, the radiative efficiency is affected by both the shock dissipation during the collision and the energy lost accelerating the post-shock gas. 

When the downstream outflow is optically thick, before photons diffuse out, the radiation force accelerates gas and converts radiation energy to kinetic energy, yielding lower net radiative efficiency. We assess this by estimating the ratio of radiation diffusion time to the advection time in the downstream gas. We select outflowing gas as described in Section~\ref{subsec:discussion_compare}. Then we calculate the average optical depth and radial velocity measured relative to the collision point to estimate $t_{\rm diff}\sim \tau R/c$ and $t_{\rm adv}\sim R/v_{\rm r}$ respectively. We list the estimated $t_{\rm diff}/t_{\rm adv}$ in Table~\ref{tab:rad_eff_fedd} together with the radiative efficiency. The sub-Eddington run  A122Edd001 shows significantly shorter diffusion time than advection time, and $t_{\rm diff}/t_{\rm adv}$ increases as mass fallback rate becomes larger. The ratio $t_{\rm diff}/t_{\rm adv}$ of A122Edd1 and A122Edd10 are well above unity, suggesting photon diffusion is slow in these two runs. Interestingly, A122Edd1 and A122Edd10 show similar radiative efficiency despite the order of magnitude different fallback rate, which may suggest that there is a minimum radiative efficiency for a fixed stream orbit.   

\begin{table}
\caption{Radiation efficiencies}
\label{tab:rad_eff_fedd}
\begin{threeparttable}
\begin{tabular}{lcccc}
\hline
$\dot{M}/\dot{M}_{\rm Edd}$ & $10^{-2}$ & $10^{-1}$ & 1.0 & 10.0\\
\hline
$\eta_{\rm rad}$ & $\sim60.3\%$ & $\sim18.5\%$ & $\sim6.2\%$ & $\sim3.8\%$\\
\hline
$t_{\rm diff}/t_{\rm adv}$ & $\sim0.4$ & $\sim2.0$ & $\sim17.3$ & $\sim133.4$\\ 
\hline
\end{tabular}
\end{threeparttable}
\end{table}

\begin{figure}
    \centering
    \includegraphics[width=\linewidth]{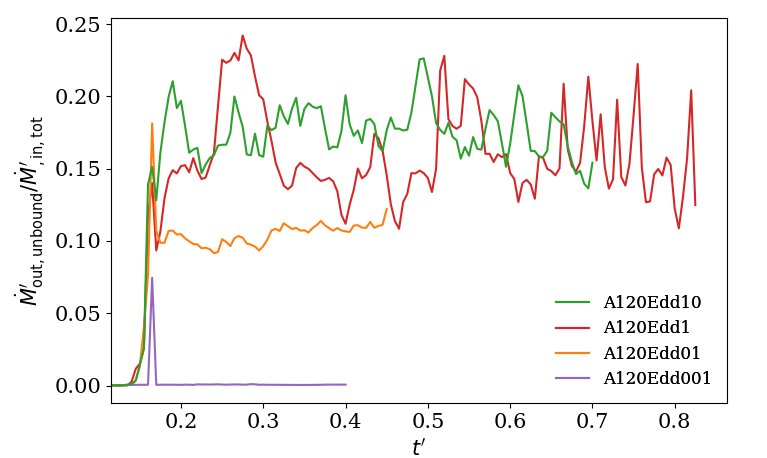}
    \caption{Total unbound mass flux as a function of time, normalized to the total injected mass flux. The fluxes are calculated at the simulation domain up to $r=40.0$( i.e. integrated over all the surface of the volume $(27,40)\times(0.42\pi,0.58\pi)\times(0.81\pi,0.96\pi)$). We also calculate mass flux at $r=45.0,~50.0$ and did not find significant difference in the quasi steady state.}
    \label{fig:mdot_massrate}
\end{figure}

The kinetic energy efficiency is shown in Figure~\ref{fig:efficiency_massrate}, including total kinetic energy $\eta_{\rm KE}$ (the solid lines) and the unbound kinetic energy $\eta_{\rm KE,ub}$ (the dashed lines). Consistent with the lower radiative efficiency, the higher mass fallback rate usually leads to higher kinetic energy efficiency. The unbound fraction seems to follow similar trend. The transient peak near $t\sim0.16$ correspond to the time that the returning stream first encounters the fallback stream and breaks the fallback stream. Part of the broken stream left the domain as marginally-unbound gas, we note that this initial transient may be sensitive to details of domain set-up, and thus mainly focus on the quasi-steady-state energy conversion.

In Figure~\ref{fig:mdot_massrate}, we show the ratio between unbound mass flux and the total injected mass flux. A122Edd001 does not produce significant unbound mass, while A122Edd01 shows that $\sim10\%$ injected mass flux becomes unbound, A122Edd1 and A122Edd10 shows $\sim15\%$ unbound fluxes. We also find that the average velocity of unbound gas are factor of a few similar in A122Edd01, A122Edd1 and A122Edd10. The results are generally consistent with the unbound fraction found in \citet{jiang2016prompt} with comparable Eddington ratio.

\subsection{Effect of collision angle}\label{subsec:result_angle}
\begin{figure}
    \centering
    \includegraphics[width=\linewidth]{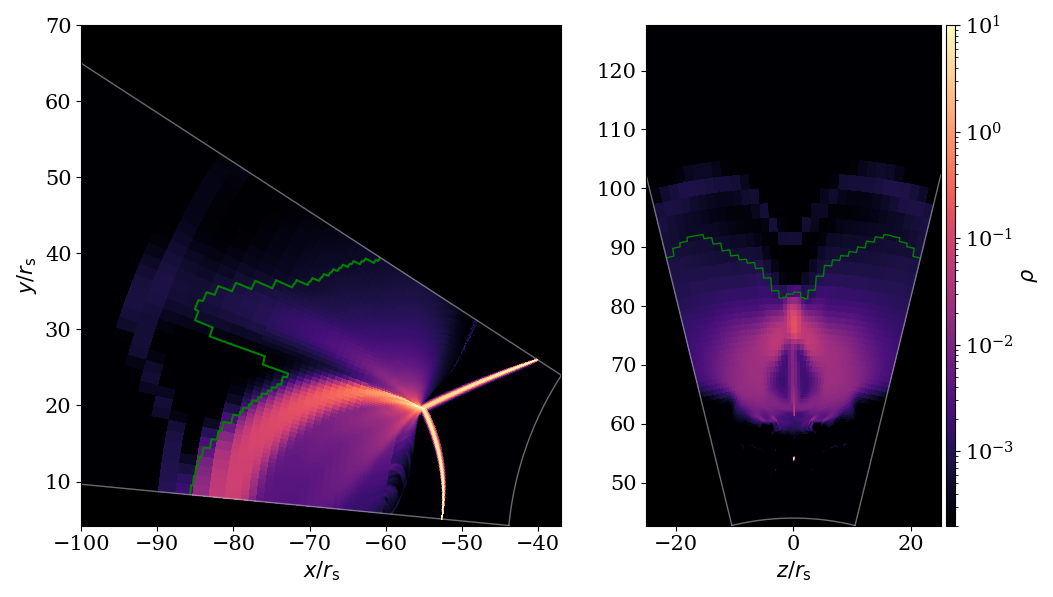}
    \caption{Gas density snapshots at $t=0.45$ for A90Edd1. The left plot is at $\theta=\pi/2$-plane, the right plot is at mid-$\phi$-plane. The green solid line shows the photosphere surface as defined in Equation~\ref{eq:tau1}.}
    \label{fig:a90_dens}
\end{figure}

In A90Edd1, we adjust the orbital parameter of the returning stream to have collision angle $\approx90^{\circ}$, while keeping the collision radius similar to A122Edd1 ($r_{\rm coll}=24.1\rs$, Table~\ref{tab:sim_params}). Figure~\ref{fig:a90_dens} shows the gas density snapshots from A90Edd1 after it has reached a quasi-steady-state. The majority of downstream gas forms a stream-structure that expands in $\theta$ direction perpendicular to the orbital plane. We find that the stream-structure is bounded and falls towards the black hole.

The collision creates unbound gas away from the orbital plane. They are spatially divided into two groups. The first group of unbound gas stays in the inner side of the merged stream, with velocity oriented towards the black hole and leaving the calculation domain from the outer $\phi$ boundary. These unbound gas is relatively diffuse, with enhanced velocity due to the decreasing gravitational potential when approaching black hole. The second group of unbound gas is ejected from the outer side of the merged stream moving outward radially, with higher density than the first group. We find that the total unbound mass flux is $\sim10\%$ of total injected mass flux, with $\sim7.6\%$ in the $r$ direction contributed by the second of unbound gas, and $\sim 2.4\%$ in the $\phi$ direction contributed by the first group of diffusing gas. 

\begin{figure}
    \centering
    \includegraphics[width=\linewidth]{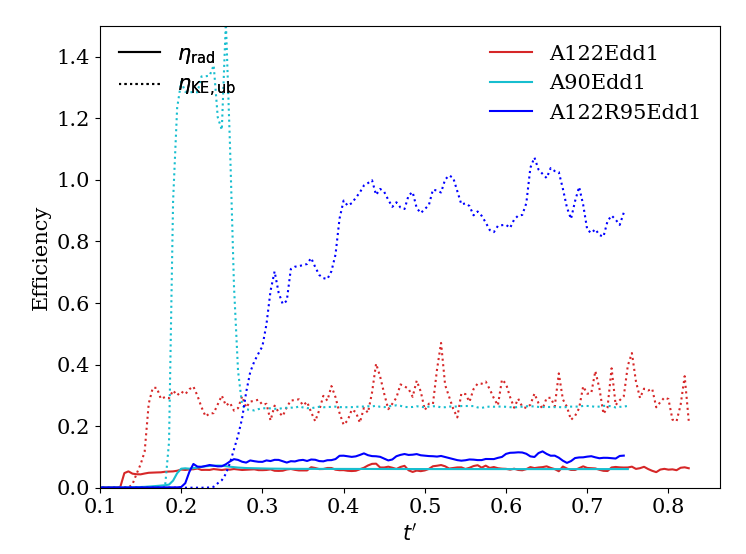}
    \caption{radiative efficiency (the solid line) and unbound kinetic energy efficiency (the dotted lines) as a function of time. The red lines are for A122Edd1, and the cyan lines are for A90Edd1, the dark blue lines are for A122R95Edd1. The kinetic energy efficiency might be larger than 1.0 because of conversion between kinetic energy and gravitational potential. In A90Edd1, there is also a transient phase ($0.17\gtrsim t'\gtrsim0.26$) with high unbound kinetic energy efficiency, which correspond to the time when the approaching stream is broke by the returning stream, the disrupted stream is marginally unbound and leaving the domain. We note that this transient high unbound rate is artifact due to our domain set-up.}
    \label{fig:efficiency_orbit}
\end{figure}

Figure~\ref{fig:efficiency_orbit} compares the unbound kinetic energy efficiency and the radiative efficiency in A90Edd1 with A122Edd1. Despite the similar total unbound kinetic energy efficiency, the contribution from $r$, $\theta$ and $\phi$ boundaries are significantly different, suggesting different downstream gas morphology. We find that the $\sim70\%$ of the unbound kinetic energy flux is from the $r$ direction, carried by the second group of massive, fast unbound gas. The rest unbound kinetic energy flux is from the first group of diffuse unbound gas with lower velocity. The $\theta$  boundary contributions are negligible. The uneven distribution is in contrast with the relatively equal amount of unbound kinetic energy flux in the $r$, $\theta$ and $\phi$ boundaries in A122Edd1. 

The radiative efficiency is also similar to A122Edd1. However, this does not imply a similar radiation energy production by the collision. We find that the maximum radiation energy at the collision shock front in A90Edd1 is about 5 times lower than A122Edd1. However, the downstream gas is more optically thin in A90Edd1. The green solid lines in Figure~\ref{fig:a90_dens} show the photon $R_{\tau=1}$ at quasi-steady state. Compared to the extended optically thick region in A122Edd1, photons can diffuse out of downstream gas in A90Edd1 more easily. 

Interestingly, the comparison between A90Edd1 and A122Edd1 shows that results deviate from a ballistic collision model. If the collision is inelastic and the collision angle is $\theta_{\rm coll}$, assuming a symmetric collision velocity vector on both side of the shock, the dissipated energy will roughly be $\propto\sin(\theta_{\rm coll}^{2}/2)$ \citep{dai2015soft}, where only the velocity in the net-momentum direction is preserved. However, A90Edd1 and A122Edd1 show a similar radiative efficiency that does not scale as $\propto\sin(\theta_{\rm coll}/2)$, suggesting that the net luminosity does not solely depend on the collision angle, but also the optical depth of downstream gas. 

\subsection{Effect of collision radius}
\label{subsec:result_radius}
In A122R95Edd1, we adjust the returning stream to make the collision radius to be $\approx95\rs$, while keep the collision angle close to A122Edd1 (Table~\ref{tab:sim_params}). The modified returning stream is on a slightly unbound orbit, with $v_{\rm tot, 1}\approx0.115c$ relative to the black hole before collision. The fallback stream is on the same orbit as A122Edd1, which is slightly bounded with $v_{\rm tot, 1}\approx0.097c$ before collision. These velocities are only about half of the velocity in A122Edd1 before collision. The set-up is designed to compare with A122Edd1 to study the effect of collision radius, the realistic stream orbit collides near this radius will likely to show smaller collision angle for a spin-less black hole \citep{dai2015soft,bonnerot2021first}.
\begin{figure}
    \centering
    \includegraphics[width=\linewidth]{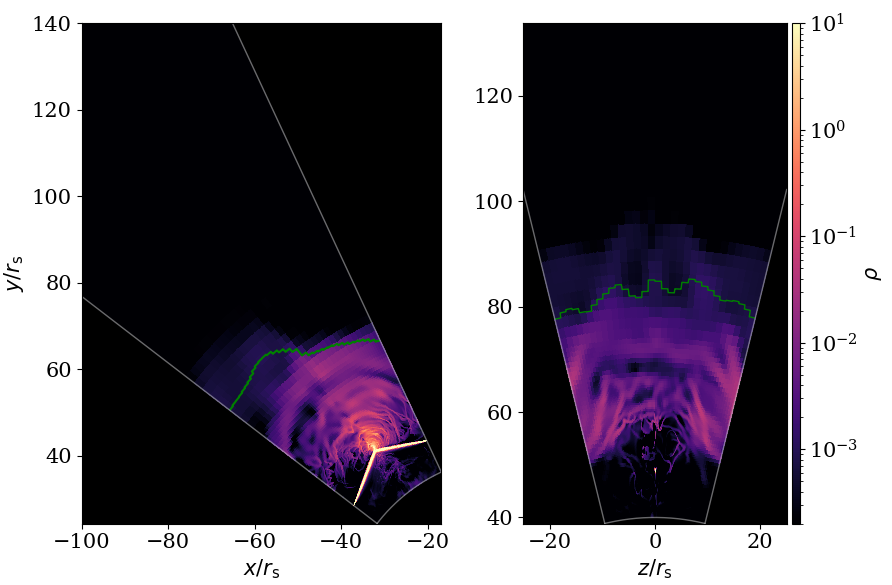}
    \caption{Gas density snapshots at $t'=0.52$ for A122R95Edd1, about $\Delta t'\sim0.34$ after the collision, which is comparable to $t'=0.45$ in A122Edd1 (the third column in Figure~\ref{fig:Result_edd1_dens}). The left plot is the $\theta=\pi/2$-plane, the right plot is the mid-$\phi$-plane. The green solid line shows the same average for the photosphere surface as defined in Equation~\ref{eq:tau1}.}
    \label{fig:r95_dens}
\end{figure}

Figure~\ref{fig:r95_dens} shows the gas density snapshot at $t'=0.52$, which is at a comparable time after collision as A122Edd1 in Figure~\ref{fig:Result_edd1_dens}. The outflow morphology is similar to A122Edd1, but the gas flow is slower and propagates to smaller radius relative to the collision point. The gas is also less bent towards the black hole due to the weaker gravity.

When moving the collision radius outward, the available kinetic energy before collision will be reduced, in our case, the speed of both streams are roughly reduced by half. So the radiation energy density at the shock will potentially be lower, suppressing the immediate acceleration of gas near the collision. However, the effect of gravity is also weaker compare to A122Edd1, so the deceleration of downstream gas is reduced. These competing effects shape the collision and  post-shock gas evolution.

Figure~\ref{fig:efficiency_orbit} shows that the radiative efficiency of A122R95Edd1 is only slightly higher than A122Edd1. We find that the gas compression rate at the shock (i.e. the ratio between stream density before and at the shock) is similar in the two simulations. The local radiation energy density in A122R95Edd1 is about five times lower, roughly consistent with the two times lower gas velocity at both sides of the shock. 

\begin{figure}
    \centering
    \includegraphics[width=\linewidth]{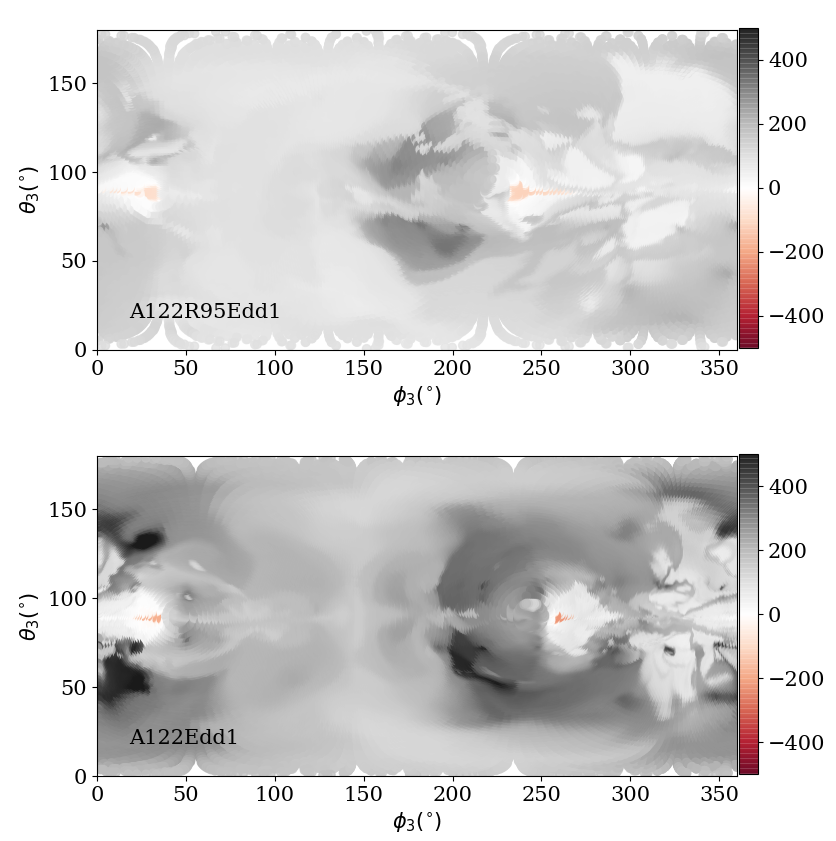}
    \caption{Angular distribution of radial velocity $v_{\rm r,3}$ relative to the collision point for all the gas that $\approx5\rs$ to the collision point. $\theta_{3}$ and $\phi_{3}$ is the polar and azimuthal angle relative to the collision, and ranges from $0-\pi$ and $0-2\pi$ respectively. The negative $v_{\rm r,3}$ corresponds to the injected streams, the positive $v_{\rm r,3}$ is normal to the sphere and pointed outward, representing the outflowing gas. The upper panel shows the distribution for A122R95Edd1 at $t'=0.52$. The lower panel shows the distribution for A122Edd1 at $t'=0.45$, roughly the same time after collision. }
    \label{fig:veltocoll}
\end{figure}

However, the unbound kinetic energy and mass flux are significantly higher than A122Edd1. We find that the unbound and injected mass flux ratio is close to 1.0 at the quasi-steady state, suggesting more massive but slower unbound outflow. Figure~\ref{fig:veltocoll} shows the angular distribution of radial velocity relative to the collision point $v_{\rm r,3}$. The outward velocities $v_{\rm r,3}>0$ in A122R95Edd1 is generally smaller than A122Edd1. We note that our definition of unbound gas is sensitive to the initial stream orbital energy. Some unbound gas could move towards the black hole and potentially interact with other gas near pericenter before escaping from the system. So the high fraction of unbound downstream gas in A122R95Edd1 should not be directly interpreted as large expanding photosphere. We will discuss the implied photosphere evolution in Section~\ref{sec:discussion}.

\section{Discussion: Stream-stream collision's role in pre-peak time}\label{sec:discussion}

\subsection{Prompt emission and contribution to luminosity}\label{subsec:discussion_promptemission}
The rise-to-peak light curve for optical TDEs seems to be more related to radiative diffusion timescale than the fallback timescale \citep{metzger2016bright, van2021seventeen}. Our simulation results roughly agree with this picture that the radiative efficiency varies with mass fallback rate. A higher fallback rate tends to produce a more optically thick outflow, which can reduce the net luminosity by reducing the diffusion rate out of the photosphere. Hence, the luminosity is not strictly proportional to the mass fallback rates. 

\begin{figure}
    \centering
    \includegraphics[width=\linewidth]{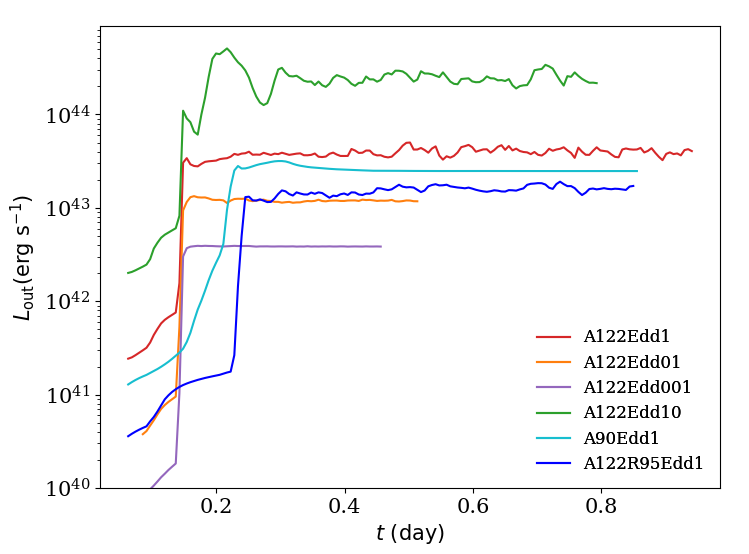}
    \caption{The total luminosity measured from $(27,40)\times(0.42\pi,0.58\pi)\times(0.81\pi,0.96\pi)$ (i.e. the simulation domain up to $r=40.0$) as a function of time. The red, orange, purple, green solid lines for mass fallback rate with $f_{\rm Edd}=0.01,0.1,1.0,10.0$ respectively. The cyan line shows the smaller collision angle of $90^{\circ}$ with $f_{\rm Edd}=1.0$. The blue line shows the larger collision radius of $r_{\rm coll}\approx95\rs$ with $f_{\rm Edd}=1.0$.}
    \label{fig:luminosity}
\end{figure}

Figure~\ref{fig:luminosity} shows the luminosities measured from the simulations. We measure the total luminosity by integrating the radiation flux on the boundary of the simulation domain $L_{\rm out}\sim\int\mathcal{F}_{\rm rad,out}dA$, with the exception that we truncate the measurement to surfaces interior to $r=40$, rather than integrating out to the outer radial boundary. We find that the measured luminosity is not sensitive to the choice of this enclosed radius as long as $r\gtrsim 35.0$. The measured luminosity includes both the diffusive and advective radiation fluxes on the $\theta$ and $\phi$ boundaries. We expect that the radiation advected through these boundaries will eventually be radiated from the photosphere, which we do not capture within the simulation domain in these directions. Hence, the timescale for this radiation to escape will be longer than estimated here, potentially leading to a lower peak luminosity but longer duration. Due to our relatively short simulation duration and the assumption of fixed mass fallback rates, our simulations can be thought of as modelling specific stages of the rising light curve. Capturing the full pre-peak luminosity rise requires consideration of time-dependent mass fallback rate in a global calculation domain that fully captures the photosphere.

Figure~\ref{fig:luminosity} shows that prior to the collision, the debris streams are the primary sources of luminosity. As stream-stream collision happens, all the simulations produce a prompt luminosity of $\sim 10^{42-44}\rm erg~s^{-1}$, making it an important pre-peak emission source. For the A122 runs, around $t=0.12~(0.13\rm day)$, the stream-stream collision enhances the luminosity by about two to three orders of magnitude. The two simulations with different orbital parameters A90Edd1 and A122R95Edd1 produce similar order-of-magnitude emission as A122Edd1. 

Although we find significant prompt emission, we infer that a one-time collision might not be sufficient to produce peak optical luminosity. This is the case even though we have adopted a favorable set-up for generating luminosity by the collision. We assume no vertical offset between the two streams, which may reduce prompt luminosity \citep{jiang2016prompt}, and we also scale the mass fallback rate to the Eddington rate for a $10^{7}M_{\odot}$ black hole. As we discuss below, we also assume the returning stream is the same width as the fallback stream, and has not spread during pericenter passage. Nevertheless, it still requires a substantially super-Eddington fallback rate to produce luminosity $\gtrsim 10^{44}\rm erg~s^{-1}$, corresponding to some observed TDE peak optical luminosity \citep{mockler2019weighing, van2021seventeen, gezari2021tidal}. For a less massive or spinning black hole, the results suggest that a highly super-Eddington fallback rate might be required to explain the peak optical luminosity by a single collision event. 

Instead, it is possible that multiple collisions or collision-like events from early-circularizing gas might be able to provide accumulative emission that is consistent with the rise-to-peak optical light curve. Another possibility is that the TDE may be interacting with a preexisting lower luminosity accretion flow \citep{chanetal2021}. The range of mass fallback rates we show in Section~\ref{subsec:result_mdot} may also represent different stages of an increasing mass fallback rate. Adding up the quasi-steady state luminosity for different $\dot{M}$ may yield increasing luminosity in the pre-peak time. In addition, when the fallback rate grows from sub-Eddington to super-Eddington, the downstream gas may change from optically-thin to optically-thick, potentially leading to spectral or luminosity variation. For a spinning black hole, the collision can be significantly delayed \citep{guillochon2015dark, hayasaki2016circularization}, so the mass fallback rate may already settle to a relatively constant value when the first encounter happens. We leave the further discussion of time-dependent mass fallback rate to future studies.

\subsection{Photosphere evolution and reprocessing layer}\label{subsec:discussion_reprocessinglayer}
For all the simulations we present (except for A122Edd001), the stream-stream collision can create an optically-thick outflow. The location of the corresponding photosphere and its evolution will depend on the line of sight. We use the photosphere radius $R_{\tau=1}$ (Equation~\ref{eq:tau1}) to estimate the scattering photosphere along the radial lines-of-sight towards collision region.

\begin{figure}
    \centering
    \includegraphics[width=\linewidth]{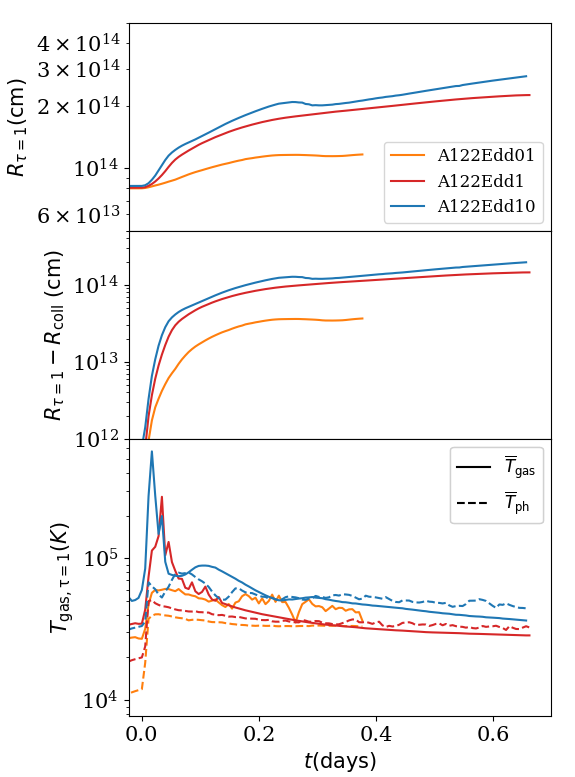}
    \caption{The evolution of average photosphere radius $R_{\tau=1}$ (the upper panel), the increment of $R_{\tau=1}$ (the middle panel) and the average temperature. The time is normalized to days since collision. In the lower panel, the solid lines are the measured average gas temperature at $R_{\tau=1}$ surface, and the dashed lines are the estimated radiation temperature according to Equation~\ref{eq:Tph}.}
    \label{fig:discussion_tauavg}
\end{figure}

Figure~\ref{fig:discussion_tauavg} shows the average $R_{\tau=1}$ for A122Edd01, A122Edd1, and A122Edd10. The collision is able to enhance the average $R_{\tau=1}$ by one to two orders of magnitude, reaching the quasi-steady state size of $\sim10^{14}\rm cm$. The super-Eddington run A122Edd10 shows a transition near $t\sim0.25$ days, where we observe that the distribution of $R_{\tau=1}$ changes from single-peak to bi-modal, and then becomes single-peak again. This corresponds to a transient dynamical stage that a leading shell of gas becomes optically-thin as it expands, and the $R_{\tau=1}$ distribution starts to be dominated by other later-accelerated gas. The sub-Eddington run A122Edd01 has a smaller $R_{\tau=1}$ and also a slower growth rate, consistent with lower radiation pressure acceleration. In the middle row of Figure~\ref{fig:discussion_tauavg}, we also plot the difference between $R_{\tau=1}$ and the collision radius $R_{\rm coll}$ to better show the photospheric expansion.

\begin{figure}
    \centering
    \includegraphics[width=\linewidth]{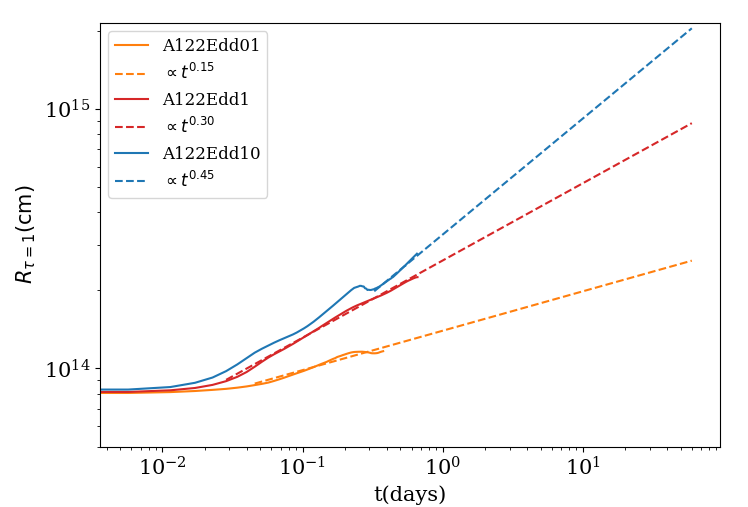}
    \caption{The extrapolated $R_{\tau=1}$ evolution for A122Edd01 (orange), A122Edd1 (red) and A122Edd10 (blue). The dashed line are the fitted power-law extrapolation, the slopes are $0.15$, $0.30$ and $0.45$ for A122Edd01, A122Edd1 and A122Edd10.}
    \label{fig:logtau}
\end{figure}

The growth $R_{\tau=1}$ roughly follows power-law for the A122 simulations. In Figure~\ref{fig:logtau}, we show the extrapolation of the $R_{\tau=1}$ evolution up to $~\sim 60$ days. However, we also note that the extrapolation should not be interpreted as predicted photosphere evolution. We will show later that, the weaker acceleration runs such as A122Edd01, A90Edd1 and A122R95Edd1 indicate that the power-law usually can describe the expansion stage, but the growth will flatten as the outflow decelerates. Properly constraining the long term gas dynamics and the inferred photosphere evolution requires global simulations. 

The solid lines in the lower panel of Figure~\ref{fig:discussion_tauavg} show the average gas temperature $\overline{T}_{\rm gas}$ measured at $R_{\tau=1}$ surface. Shortly after the collision, the hot post-shock gas shows an average temperature $\sim10^{5}$K. As the outflow expands and cools, the average gas temperature asymptotes to a few $10^{4}$K, consistent with commonly inferred blackbody temperature in optical/UV TDEs \citep{gezari2021tidal}. 

We also define a photosphere temperature $\overline{T}_{\rm ph}$ based on $R_{\tau=1}$ and the total luminosity $L_{\rm tot}$ we measured in simulations (Figure~\ref{fig:luminosity}) by:
\begin{equation}\label{eq:Tph}
    \overline{T}_{\rm ph}=\left(\frac{L_{\rm tot}}{4\pi\sigma_{\rm SB} R_{\tau=1}^{2}}\right)^{1/4}
\end{equation}
We plot $\overline{T}_{\rm ph}$ in Figure~\ref{fig:discussion_tauavg} as the dashed lines. They agree with average gas temperature  $\overline{T}_{\rm gas}$ well at late time when the photosphere becomes more isotropic, but deviate from $\overline{T}_{\rm gas}$ at early time because of the anisotropy of photosphere.

\begin{figure}
    \centering
    \includegraphics[width=\linewidth]{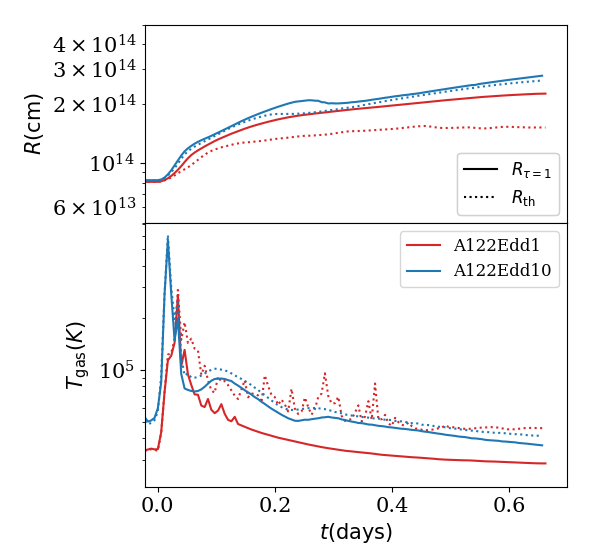}
    \caption{Upper panel: Comparing average scatter photosphere radius $R_{\tau=1}$ (the solid lines) and thermalization radius $R_{\rm th}$ (the dotted lines). The time is normalized to days since collision. Lower panel: the measured average gas temperature at $R_{\tau=1}$ (the solid lines) surface and $R_{\rm th}$ (the dotted lines) surface.}
    \label{fig:rth_avg}
\end{figure}

We also estimated a thermalization radius $R_{\rm th}$, which roughly characterizes the surface exterior to which a typical photon can be emitted and leave the domain without be absorbed \citep{1986rpa..book.....R}. Here we adopt a definition similar to $R_{\tau=1}$, but replace $\kappas$ by $\sqrt{\kappas\kappaa}$ (assuming $\kappas \gg \kappaa$) in Equation~\ref{eq:tau1}. We found that the average $R_{\rm th}$ is generally smaller than $R_{\tau=1}$. For example, Figure~\ref{fig:rth_avg} compares average $R_{\rm th}$ and $R_{\tau=1}$ in A122Edd1 and A122Edd10, where both quantities are relatively uniform. In both cases, $R_{\tau=1}$ is slightly larger than $R_{\rm th}$. The downstream gas in A122Edd10 is more optically thick than A122Edd1, and the two radii track each other more closely and the average gas temperature shows less difference. In A122Edd001 and A122Edd01, the location of $R_{\rm th}$ is highly anisotropic about the collision point due to the downstream gas flow being more optically thin. Nevertheless, the results are qualitatively consistent with the higher fallback rate runs in that its location is interior to the photosphere and the temperature are larger than those at the photosphere.

\begin{figure}
    \centering
    \includegraphics[width=\linewidth]{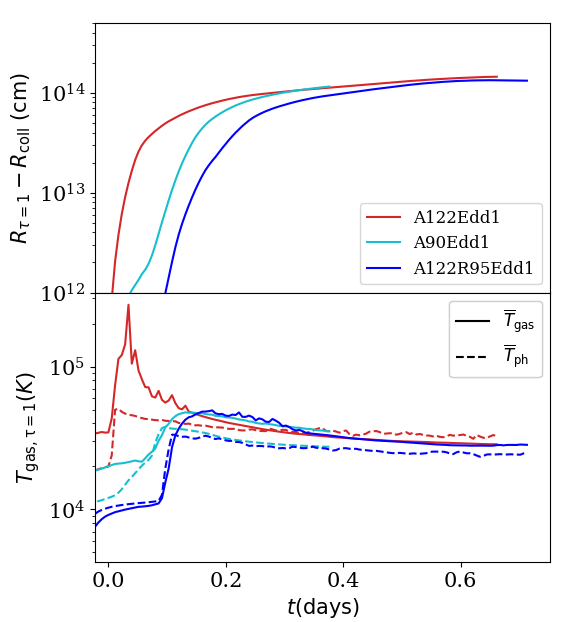}
    \caption{The evolution of average photosphere radius $R_{\tau=1}$ (the upper panel) and the average temperature for A122Edd1 (red), A90Edd1 (cyan) and A122R95Edd1 (blue), plotted time is normalized to days since the collision. In the lower panel, the solid lines are average gas temperature at $R_{\tau=1}$ surface, the dashed lines are the estimated photosphere temperature according to Equation~\ref{eq:Tph}}
    \label{fig:tauavg_orbit}
\end{figure}

Figure~\ref{fig:tauavg_orbit} shows $R_{\tau=1}$ for the two runs with modified orbits. The general evolution is similar to the A122Edd1. In A122R95Edd1, evolution of $R_{\tau=1}$ flattens more towards the end of simulation when compare to A122Edd1. But the the estimated photospheric temperature $\overline{T}_{\rm ph}$ is usually lower than $\overline{T}_{\rm gas}$ in quasi-steady state. We find that the average gas temperature at $R_{\tau=1}$ is relatively uniform, the discrepancy is more relevant to the inhomogeneous shape of photosphere. For A122Edd1 and A122Edd10, the angular distribution of $R_{\tau=1}$ and $\overline{T}_{\rm gas}$ is relatively uniform. In contrast, the angular distribution of $R_{\tau=1}$ show larger dispersion in A122Edd001, A122Edd01 and A90Edd1, suggesting that the photosphere is more ellipsoidal than spherical. For example, A122Edd1 has average photosphere radius $R_{\tau=1}\approx67.10\rs$ with dispersion $\delta_{R_{\tau=1}}\approx2.94\rs$ at $t=0.45$, but A90Edd1 has $R_{\tau=1}\approx78.68\rs$ with dispersion $\delta_{R_{\tau=1}}\approx12.25\rs$ at $t=0.45$. Using the spherical assumption can overestimate the surface area and underestimate $\overline{T}_{\rm ph}$, but it normally does not lead to an order-of-magnitude difference. 

\subsection{Comparison to Previous Studies}\label{subsec:discussion_compare}
\textbf{Angular Momentum of Downstream Gas}: 
In our simulations, the strong radiation pressure produced in the stream-stream collision sufficiently redistributes downstream gas angular momentum. \citet{lu2020self} show that the collision is able to cause a significant spread in downstream gas angular momentum with respect to the original stellar orbital plane, enabling the formation of an extended accretion disk. Our findings also agree with this picture. Following \citet{tejeda2013accurate}, we define total angular momentum as $|l|=r_{1}^{3}(\dot{\theta}_{1}^{2}+\rm sin^{2}_{1}\dot{\phi}_{1}^{2})^{1/2}/(r_{1}-\rs)$, and orbital-plane angular momentum as $l_{\rm z}=r_{1}^{3}\dot{\phi}_{1}/(r_{1}-\rs)$, where the velocities and radius corresponds to the black hole. 

\begin{figure}
    \centering
    \includegraphics[width=\linewidth]{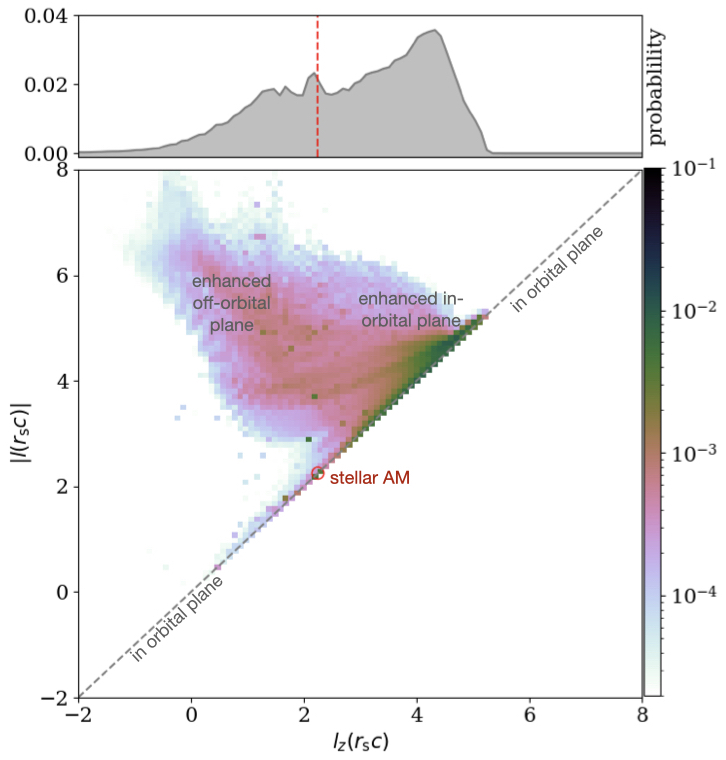}
    \caption{Downstream gas angular momentum distribution from A122Edd1 at t=0.3. We only show gas with positive radial velocity relative to the collision point $v_{r,3}>0$. The first row shows the (mass weighted) angular momentum $l_{\rm z}$ distribution projected to the orbital plane. The red vertical line labels the original stellar debris stream angular momentum. The second row shows the distribution of total angular momentum magnitude $|l|$ and projected angular momentum $l_{\rm z}$. The red circle shows the original stellar stream value, the gray dashed line labels the angular momentum within the orbital plane.}
    \label{fig:amamhist}
\end{figure}

Figure~\ref{fig:amamhist} shows the distribution of outflow angular momentum, where we select gas moving away from the collision point. The first row shows that orbital-plane angular momentum $l_{\rm z}$ is significantly spread by the collision. Compared to \citet{lu2020self}, we find the distribution is more asymmetric relative to the initial stellar angular momentum and the retrograde fraction (negative $l_{\rm z}$) is lower, potentially due to stronger radiative acceleration in the radial direction.  

The second row of Figure~\ref{fig:amamhist} shows the distribution of $|l|$ and $l_{\rm z}$. The gray dashed line labels the orbital plane, where $|l|=l_{\rm z}$. In the orbital plane, a noticeable fraction of gas gains angular momentum. Off the orbital plane, some gas shows low $l_{\rm z}$ but enhanced $|l|$, we also find this gas is usually unbound, corresponding to the unbound outflow propagating away from the orbital plane. A122Edd10 shows a similar distribution, with a slightly stronger off-orbital plane outflow. Other simulations also show a significant spread in the specific angular momentum distribution, but the relative distribution of $|l|$ and $l_{\rm z}$ is sensitive to the outflow morphology. For example, A90Edd1 and A122R95Edd1 both show weaker off-orbital plane outflow, so a tighter distribution in $|l|$. 

Previous studies had proposed a quasi-spherical envelope or wind from the early-circularizing gas to explain the reprocessed optical emission \citep{ulmer1999flares, roth2016x, metzger2022cooling}. The off-orbital plane outflow with enhanced angular momentum we observed in A122Edd1 and A122Edd10 may contribute to the formation of such pressure-supported envelope at early times, but the long-term evolution of these outflow gas should be tracked in an extended calculation domain to confirm.

\textbf{Outflow morphology}: 
Some previous models for optical emission include a quasi-spherical wind near the disk. An important difference of the outflow observed in our simulations is its inhomogeneous, asymmetric morphology. Compared to \citet{jiang2016prompt}, the inclusion of gravity changes the outflow morphology at larger radius, especially when the collision angle is small, such as in A90Edd1. \citet{lu2020self} tracks the streamlines of collision induced outflow elements, and estimates that the outflow gas distributes roughly within a cone-geometry. We also find that the outflow produced in A122Edd1, A122Edd10 and A122R95Edd1 is roughly within a cone centered on the collision point, but generally less oriented towards the black hole, potentially due to the redistribution by radiation pressure.

A commonly used assumption of the outflow/wind is quasi-spherical morphology with relatively constant radial velocity and $\propto r^{-\xi}$ density profile \citep{strubbe2009optical, jiang2016prompt, bonnerot2021first}. For example, assume a spherical outflow the with density $\rho\propto r^{-2}$. Within the photon-trapping radius, the outflow expands nearly adiabatically, the radiation energy density $U_{\rm r}\propto P_{\rm r}\propto \rho^{4/3}\propto r^{-8/3}$ \citep{strubbe2009optical}, so the luminosity at a radius roughly follows $L\propto 4\pi r^{2}U_{\rm r}\propto r^{-2/3}$. 

In A122Edd1, A122Edd10 and A122R95Edd1, we find that only within a cone of $\sim 45^{\circ}$ with distance of $\lesssim15\rs$ relative to the collision point, the average radial velocity is roughly constant ($\sim 0.1c$) and the density distribution roughly follows $\propto r^{-2}$. For example, at $t=0.3$, A122Edd1 show an average density profile $\propto d_{\rm coll}^{-2.3}$, where $d_{\rm coll}$ is the distance to collision point. We also find that within the cone, the radiation energy density profile $\sim d_{\rm coll}^{-3.2}$, which is a slightly steeper drop than $\sim r^{-8/3}$, potentially due to the diffusive radiation flux. Outside the cone, the velocity distribution is anisotropic, especially near the streams. For example, the lower panel of Figure~\ref{fig:veltocoll} shows the radial velocity distribution of A122Edd1 on a sphere that is $5\rs$ from the collision point. The velocity distribution within $100^{\circ}\lesssim\phi_{3}\lesssim190^{\circ}$, $v_{\rm r,3}$ is relatively uniform, but it is anisotropic in other regions. In other simulations, the outflow morphology is even more anisotropic.

Such asymmetric outflow morphology may affect the radiative efficiency too. Compared to a purely adiabatic spherical outflow, radiation may preferentially diffuse out from the optically thin regions near the collision point during expansion. For simplicity, we can assume all the stream kinetic energy is deposited as luminosity $L_{\rm coll}$ immediately at collision within a spherical region of $r\sim H$, where $H$ is the width of the stream. Assuming no diffusion and adopting the $L \propto r^{-2/3}$ dependence derived above,  radiative efficiency can be roughly estimated as the ratio between $L_{\rm coll}$ and the luminosity escaping at $r_{\tau=1}$ , so $\eta_{\rm rad}\sim(r_{\tau=1}/H)^{-2/3}\sim((R_{\tau=1}/10^{14}\rm cm)/(H/3R_{\odot}))^{-2/3}\sim1.6\%$. 

The radiative efficiency $\eta_{\rm rad}$ we found is usually higher than this estimation, consistent with enhanced radiative losses due to photon diffusion. Similar to the total luminosity $L_{\rm out}\sim\int\mathcal{F}_{\rm rad,out}\cdot \mathbf{dA}$ (Figure~\ref{fig:luminosity}), we can estimate the ``advective'' part by replacing $\mathcal{F}_{\rm rad,out}$ by $\mathcal{F}_{\rm rad,adv}\sim \mathbf{v}E_{\rm rad}$, where $\mathbf{v}$ is gas velocity. So the difference between total luminosity and the advection part provides an estimate of the diffusive part. We find the advective fraction increases as $f_{\rm edd}$ increases, consistent with previous speculation. But we also find even for higher mass flux runs, there is a non-negligible diffusive fraction. For example, for A122Edd1, the advection fraction is $\sim 80\%$ from the optically thick region near to the collision point, and drops to $\sim68\%$ as gas moves outward. The total radiation flux from the $\phi$ direction boundaries has two to three times lower advection fraction than the other boundaries. These findings suggest the expansion of the outflow is not strictly adiabatic due to the photon diffusion within the anisotropic outflow.

\textbf{Returning Stream Width}: We assume the returning stream has the same width as the fallback stream in the simulations. Previous studies of TDEs with low mass ratios \citep{ramirez2009star, rosswog2009tidal, shiokawa2015general} or deep penetration encounters \citep{sadowski2016magnetohydrodynamical} suggest thickened streams due to the strong nozzle shock near pericenter. Recent studies of TDEs by more massive black hole can also show significantly expansion in the returning stream, potentially changing the collision initial condition and outcome \citep{liptai2019disc,curd2021global, andalman2022tidal, steinberg2022origins}. 

\citet{bonnerot2021formation} compares the potential outcomes of stream-stream collision with different returning stream widths. They suggest that a thickened returning stream with reduced density could sweep through the thin stream without significant deflection, suppressing the radiation pressure produced during stream-crossing, making it less likely to drive strong outflow. \citet{steinberg2022origins} adopt a equation of state including hydrogen recombination and suggests the recombination heating near pericenter can significantly expand the stream and potentially enhance the nozzle shock dissipation. \citet{coughlin2023dynamics} also suggest that the steam density can be lowered by the recombination processes near pericenter. \citet{bonnerot2022nozzle} performed high resolution study of the vertical compression of the stream and energy dissipation due to the nozzle shock. They, however, suggest that the stream is likely to stay relatively thin after the pericenter, and that the dissipation is sensitive to compression. The stream could be compressed to $\sim 0.1\%$ of the stellar radius in vertical direction and $\sim10\%$ in the orbital plane. Such large compression rate requires sufficient resolution to minimize numerical dissipation and capture the stream expansion accurately after pericenter. We also note that our assumption of the same stream width at in-plane and vertical direction is idealized. Recent works \citep{bonnerot2022pericenter, coughlin2023dynamics} suggest that the stream geometry is set by the interplay between tidal force, pressure support and self-gravity. The relative importance of these forces is likely sensitive to gas density. As the gas orbits around the black hole, the stream cross-section can deviate significantly from circular. 

In the global test calculations with static mesh refinement (similar set-up to Figure~\ref{fig:appendix_globalorbit}, not presented in this paper), we find qualitative consistent results. Without radiation, the hydro dynamical simulations do not show convergence to the returning stream width at our highest test resolution ($\delta R\sim0.16\rs$). We find generally less spreading of the stream after pericenter as we increase the resolution. Interestingly, when we include radiation, the RHD test simulations (similar to Figure~\ref{fig:appendix_globalorbit}, with same or lower resolution ($\delta R\gtrsim0.16\rs$), not presented in this paper), we find the stream stays almost the same width after the pericenter regardless of resolution, but this may be due to insufficient resolution in the test calculations, leading to an overestimation of radiative cooling. When the thinnest part of the stream at the pericenter is only resolved by a few cells, the radiative cooling can be overestimated, reducing internal energy dissipated in the nozzle shock. Resolving the stream near the pericenter is essential to understanding the role of stream-stream collision as pre-peak emission mechanisms of TDEs. We plan to explore the collision between streams with different widths in the future work.

\section{Conclusions}\label{sec:conclusion}
We study the stream-stream collision in TDEs through a series of three-dimensional RHD simulations. We adopt a stream orbit based on a $10^{7}M_{\odot}$ black hole and a solar type star, with the penetration factor of $\beta=1.0$. We compare the effect of different mass fall back rate ($f_{\rm edd}=0.01, 0.1, 1.0, 10.0$), the collision angle ($\theta_{\rm coll}\approx122^{\circ}, 90^{\circ}$), and collision radius ($r_{\rm coll}\approx24\rs, 95\rs$). We study the radiation pressure driven outflow and potential optical emission from the stream-stream collision. We summarize our findings as follows:

\textbf{(1) Dependency on Resolution}: We find that RHD simulations of stream-stream collisions are very sensitive to resolution. Insufficient resolution may artificially thicken the stream, reducing the stream vertical optical depth and the kinetic energy density before collision. These effects suppress radiation energy generation at the collision, leading to a qualitatively inaccurate picture of downstream gas distribution (Figure~\ref{fig:result_resolution_density}). The under-resolved simulations might also underestimate the outflow optical-depth, allowing radiation to escape before it can interact sufficiently with gas. In our calculations, we use spatial resolution of $\sim 10^{-2}\rs$ (or $\sim0.85R_{\odot}$). We find that minimum $48$ angular bins for radiation is required in order to reach convergence in the properties of the post collision flow. 

\textbf{(2) radiative efficiency and Luminosity}: With the assumption of thin streams without vertical off-set, the stream-stream collision can convert $\gtrsim5\%$ of kinetic energy to radiation, producing a luminosity of $\sim10^{43-45}\rm erg~s^{-1}$. The radiative efficiency $\eta_{\rm rad}$ is affected by both the energy conversion at the collision and the interaction between radiation and downstream gas.

When the stream collision angle and radius is fixed, the radiative efficiency $\eta_{\rm rad}$ is sensitive to the mass fallback rates (Figure~\ref{fig:efficiency_massrate}) and could range from $\sim 5\%-60\%$. The higher fallback rates usually show lower radiative efficiency. The radiation pressures produced in the collision are usually higher, but drive optically-thick outflows. In these outflows, radiation accelerates gas, converting radiation energy back to gas kinetic energy. In contrast, lower mass fallback rates usually yield more optically-thin downstream gas, where radiation diffuses out more easily, leading to higher luminosity and net radiative efficiency.

When the mass fallback rate and collision radius are fixed, the smaller collision angle usually produces less radiation from the collision, resulting in lower radiation pressure. This can lead to more optically-thin downstream gas, where radiation diffusion is enhanced.  Due to the competing effects from the kinetic energy dissipation at the collision and downstream gas optical depth, A122Edd1 and A90Edd1 yield similar radiative efficiency despite the different collision angle. The radiative efficiency seems less sensitive to the collision radius. A122Edd1 and A122R95Edd1 shows similar shock structure at the collision and downstream gas distribution, resulting similar radiative efficiency. 

\textbf{(3) Radiation Driven Outflow}: In the simulations, when the mass fallback rate is $\gtrsim10\%$ Eddington, the radiation pressure can accelerates gas to form expanding optically-thick outflow, making $\sim10\%-15\%$ mass of gas unbound. The unbound gas is preferentially propagating off the orbital plane, with an order-of-magnitude $\sim 0.1c$ radial velocity relative to the collision point.

The collision largely redistributes downstream gas specific angular momentum, resulting highly aspherical outflow. In A122Edd1, A122Edd10, A122R95Edd1, the outflow has an approximately conical geometry. Within the cone, the density profile roughly follows $\propto r^{-2}$, and gas moves radially away from the collision point with $\sim 0.1c$ velocity.  In A90Edd1 and A122Edd001 show more anisotropic morphology of the outflow, where a majority of gas is within a stream-structure bent towards the black hole. The outflow morphology seems to show a potential trend from stream-structure to conical-geometry as radiation pressure increases. In all simulations, the anisotropy of downstream gas leads to non-negligible radiative diffusion, and the outflow expansion is not strictly adiabatic.

\textbf{(4) Photosphere Evolution}: We estimate photosphere along the line-of-sight towards the collision region. We find that the stream-stream collision is able to expand $R_{\tau=1}$ by one or two orders of magnitude, resulting $\sim10^{14}\rm cm$ photosphere. The average gas temperature $\overline{T}_{\rm gas}$ at $R_{\tau=1}$ is usually a few $10^{4}$K. The estimated photosphere temperature $\overline{T}_{\rm ph}$ is generally consistent with $\overline{T}_{\rm gas}$ (Figure~\ref{fig:discussion_tauavg}, Figure~\ref{fig:tauavg_orbit}). These findings suggest that the stream-stream collision provides a promising mechanism to drive pre-peak photospheric evolution, and produces photosphere size consistent with optical observations.

The anisotropic outflows formed in our simulations may lead to inclination-dependent photospheric properties. For example, downstream gas could be noticeably more optically-thin outside the cone-like outflow in some runs. In other simulations, the outflow is also usually preferentially launched in some directions instead of expanding spherically relative to the collision point. 

\textbf{\textit{Caveats}} Our local calculation domain is optimized for resolving the collision shock and tracking gas motion in the net momentum direction, but the long-term outflow evolution and photosphere in $\theta$ and $\phi$ directions can be beyond the calculation domain. Studying the system in a global domain will be the focus of our future work. When modeling the streams, we ignored the effect of self-gravity \citep{guillochon2014ps1, coughlin2016structure}, the debris energy distribution \citep{spaulding2021effect}, or magnetic fields \citep{guillochon2017simulations, bonnerot2017magnetic} on the stream structure. When modeling the black hole, we adopt an approximate gravity description best suited to a spin-less black hole. The spin can significantly change the stream orbit, resulting vertical offset between the streams and potential modifications to returning stream width \citep{guillochon2015dark, hayasaki2016circularization, batra2023general}. \citet{jiang2016prompt} showed that the luminosity produced by the stream-stream collisions will be suppressed by the vertical offset if both stream are thin. \citet{jankovic2023spininduced} explored the spin-induced collision angle modification and vertical offset, and found aspherical downstream gas can also be launched due to offsetting collision, potentially leading to similar enhanced radiation diffusion in postshock gas. We also fix the mass injection rate in the simulation, but we show that the outflow optical depth and potential photosphere property can change with mass fallback rate. Including the time-dependence of the mass fallback rate and potential pre-peak emission variability is also an interesting focus for future work.

\begin{acknowledgments}
We would like to thank Enrico Ramirez-Ruiz, Morgan MacLeod, Jane Lixin Dai, Martin Pessah, Cl\'{e}mont Bonnerot, Erin Kara, Brian Metzger and Wenbin Lu for stimulating discussions and helpful insights. This work used the computational resources provided by the Advanced Research Computing Services (ARCS) at the University of Virginia. We also used the Extreme Science and Engineering Discovery Environment (XSEDE), which is supported by National Science Foundation (NSF) grant No. ACI-1053575. This work was supported by NASA Astrophysics Theory Program grant 80NSSC18K1018 and NASA Theoretical and Computational Astrophysics Networks grant GR12611. Resources supporting this work were also provided by the NASA High-End Computing (HEC) Program through the NASA Advanced Supercomputing (NAS) Division at Ames Research Center. The Center for Computational Astrophysics at the Flatiron Institute is supported by the Simons Foundation. Part of this work was carried out during the KITP program "Bridging the Gap: Accretion and Orbital Evolution in Stellar and Black Hole Binaries" and is thus supported in part by the National Science Foundation under Grant No. NSF PHY-1748958.
\end{acknowledgments}

\appendix

\section{Gravitational Acceleration}\label{appendix:grav}
We note the variables in the coordinates relative to the black hole with subscript 1 (the black hole centered frame). According to \citet{tejeda2013accurate}, the $\ddot{r}_{1},~\ddot{\theta}_{1},~\ddot{\phi}_{1}$ and gravitational accelerations in the $r_{1},~\theta_{1},~\phi_{1}$ directions are:

\begin{eqnarray}
    &a_{r,1}=-\frac{GM_{\rm BH}}{r_{1}^{2}}\left(1-\frac{\rs}{r_{1}}\right)^{2}+\frac{\rs\dot{r}_{1}^{2}}{r_{1}(r_{1}-\rs)} - \frac{3}{2}\rs(\dot{\theta}_{1}^{2}+\sin^{2}\theta_{1}\dot{\phi}_{1}^{2}),\quad a_{\theta,1}=\dot{r}_{1}\dot{\theta}_{1}\frac{\rs}{r_{1}-\rs},\quad a_{\phi,1}=\dot{r}_{1}\dot{\phi}_{1}\sin\theta_{1}\frac{\rs}{r_{1}-\rs}\label{eq:rosswogacc}\\
    &\ddot{r}_{1}=a_{r,1},\quad \ddot{\theta}_{1}=-\frac{2\dot{r}_{1}\dot{\theta}_{1}}{r_{1}}\left(\frac{r_{1}-3/2\rs}{r_{1}-\rs}\right)+\sin\theta_{1}\cos\theta_{1}\dot{\phi}_{1}^{2},\quad \ddot{\phi}_{1}=-\frac{2\dot{r}_{1}\dot{\theta}_{1}}{r_{1}}\left(\frac{r_{1}-3/2\rs}{r_{1}-\rs}\right)-2\frac{\dot{\phi}_{1}\dot{\theta}_{1}}{\tan\theta_{1}}\label{eq:rosswog}
\end{eqnarray}
where $G$ is gravitational constant and $M_{\rm BH}$ is the mass of black hole. $r_{1},~\theta_{1},~\phi_{1}$ are the coordinate with respect to the black hole, $\dot{r}_{1},~\dot{\theta}_{1},~\dot{\phi}_{1}$ are the corresponding  first derivatives, $\ddot{r}_{1},~\ddot{\theta}_{1},~\ddot{\phi}_{1}$ are the corresponding second derivatives .

To calculate the gravitational accelerations relative to the origin, first we obtain the coordinates $r,~\theta,~\phi$ and velocities $\vr=\dot{r},~\vth=r\dot{\theta},~\vph=r\sin\theta\dot{\phi}$, then transform them into Cartesian coordinates by 
\begin{eqnarray}
    &x=r \sin \theta  \cos \phi,\quad y =r\sin \theta  \sin \phi,\quad z=r \cos \theta \\
    &\dot{x}=\dot{r}\sin \theta  \cos \phi  +r\dot{\theta} \cos \theta  \cos \phi  - r\dot{\phi} \sin \theta  \sin \phi\\
    &\dot{y}=\dot{r}\sin \theta  \sin \phi  + r\dot{\theta} \cos \theta  \sin \phi + r\dot{\phi} \sin \theta  \cos \phi   \\
    &\dot{z}=\dot{r}\cos \theta  -r\dot{\theta} \sin \theta  
\end{eqnarray}
Next, we transform the positions and velocities to the black hole centered frame by:
\begin{eqnarray}
    &x_{1} = x + L_{12},\quad y_{1}=y,\quad z_{1}=z\\
    &\dot{x}_{1}=\dot{x},\quad \dot{y}_{1}=\dot{y},\quad \dot{z}_{1}=\dot{z}
\end{eqnarray}
From the Cartesian coordinate positions and velocities, the spherical polar coordinate positions and velocities $\dot{r}_{1},\quad\dot{\theta}_{1},\quad\dot{\phi}_{1}$ are:
\begin{eqnarray}
&r_{1}=\sqrt{x_{1}^{2}+y_{1}^{2}+z_{1}^{2}},\quad \theta_{1}=\cos^{-1}(z_{1}/r_{1}),\quad \phi_{1}=\tan^{-1}(y_{1}/x_{1})\label{eq:r1theta1phi1}\\
&\dot{r}_{1}=\frac{x \dot{x}_{1}+y \dot{y}_{1}+z \dot{z}_{1}}{r_{1}},\quad \dot{\theta}_{1}=\frac{-\left(x_{1}^2+y_{1}^2\right) \dot{z}_{1}+z_{1}(x_{1}\dot{x}_{1}+y_{1}\dot{y}_{1})}{r_{1}^{2}\sqrt{x_{1}^2+y_{1}^2}},\quad \dot{\phi}_{1}=\frac{x_{1} \dot{y}_{1}-y_{1} \dot{x}_{1}}{x_{1}^2+y_{1}^2}\label{eq:dr1dth1dph1}
\end{eqnarray}
Inserting Equation~\ref{eq:r1theta1phi1} and Equation~\ref{eq:dr1dth1dph1} into Equation~\ref{eq:rosswogacc} and Equation~\ref{eq:rosswog}, we can get the accelerations and coordinate derivatives in the black hole centered frame. The acceleration in the black hole centered frame and the accelerations in our domain (relative to the origin) are the same in Cartesian coordinates, so 
\begin{eqnarray}
    &\ddot{x}=\ddot{x}_{1}=-r_{1}\ddot{\phi}_{1} \sin \theta_{1} \sin \phi_{1} +r_{1}\ddot{\theta}_{1} \cos \theta_{1} \cos \phi_{1} +
   \ddot{r}_{1}\sin \theta_{1} \cos \phi_{1}\nonumber\\
   &-2\dot{r}_{1} \dot{\phi}_{1} \sin \theta_{1} \sin \phi_{1} +2\dot{r}_{1} \dot{\theta}_{1} \cos \theta_{1} \cos \phi_{1} -2 r\dot{\theta}_{1} \dot{\phi}_{1} \cos \theta_{1} \sin \phi_{1}\nonumber\\
   &-r\dot{\theta}_{1}^2 \sin \theta_{1} \cos \phi_{1} -r\dot{\phi}_{1}^2\sin \theta_{1} \cos \phi_{1}\\
   &\ddot{y}=\ddot{y}_{1}=\ddot{r}_{1}\sin \theta_{1} \sin \phi_{1} +r\ddot{\phi}_{1} \sin \theta_{1} \cos \phi_{1} +r\ddot{\theta}_{1} \cos \theta_{1} \sin \phi_{1} \nonumber\\
   &+2 r\dot{\theta}_{1} \dot{\phi}_{1} \cos \theta_{1} \cos \phi_{1} +2\dot{r}_{1} \dot{\theta}_{1} \cos \theta_{1} \sin \phi_{1} +2\dot{r}_{1} \dot{\phi}_{1}  \sin \theta_{1} \cos \phi_{1}\nonumber\\
    &-r\dot{\theta}_{1}^2 \sin \theta_{1} \sin \phi_{1} -r\dot{\phi}_{1}^2 \sin \theta_{1} \sin \phi_{1} \\
   &\ddot{z}=\ddot{z}_{1}=-r\ddot{\theta}_{1} \sin \theta_{1}+\ddot{r}_{1}\cos \theta_{1} -2\dot{r}_{1} \dot{\theta}_{1} \sin \theta_{1} -r\dot{\theta}_{1}^2 \cos \theta_{1}
\end{eqnarray}
Accordingly, the accelerations in $r,~\theta~\phi$ directions can be obtained from the Cartesian accelerations 
\begin{eqnarray}
    &a_{r} = \ddot{x}\sin\theta\cos\phi + \ddot{y}\sin\theta\sin\phi + \ddot{z}\cos\theta\\
    &a_{\theta} = \ddot{x}\cos\theta\cos\phi + \ddot{y}\cos\theta\sin\phi - \ddot{z}\sin\theta\\
    &a_{\phi} = -\ddot{x}\sin\phi + \ddot{y}\cos\phi
\end{eqnarray}
In Equation~\ref{eq:gasmom}, \ref{eq:gasenergy}, the gravitational acceleration $\textbf{a}_{\rm grav}$ refers to $\textbf{a}_{\rm grav}=\left(a_{r}, a_{\theta}, a_{\phi}\right)$.

\begin{figure}
    \centering
    \includegraphics[width=0.5\textwidth]{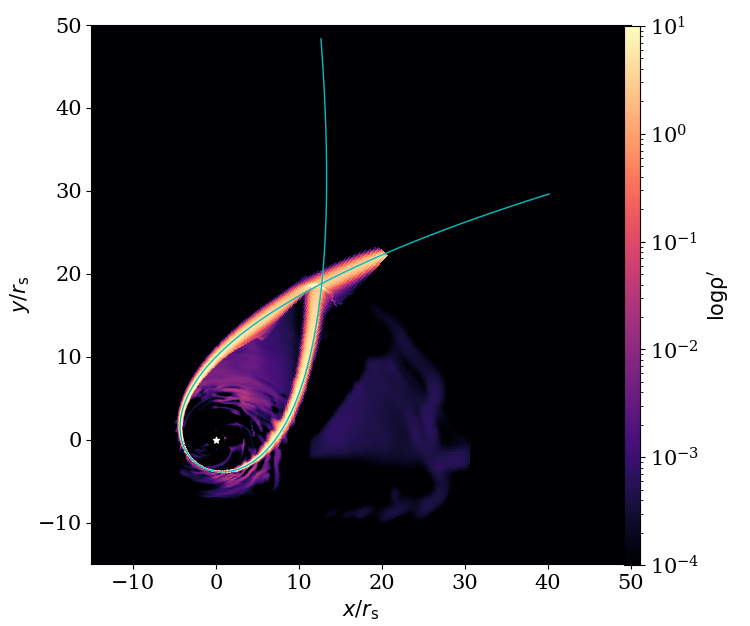}
    \caption{$\theta=\pi/2$-plane gas density snapshot of A122Edd1 at $t'=40$, overlapped with the ballistic trajectory we adopt (the cyan solid line, see also Figure~\ref{fig:method_domain}): $M_{\rm BH}=10^{7}M_{\odot},~M_{*}=M_{\odot},~R_{*}=R_{\odot},~\beta=1.0$. The good match between the streams before collision and ballistic orbit validates our gravity implementation. The white star marks the black hole location.}
    \label{fig:appendix_globalorbit}
\end{figure}

In order to verify the gravity implementation, we compared the stream trajectory before stream-stream collision with the ballistic trajectory integrated according to Equation~\ref{eq:rosswog} in Figure~\ref{fig:appendix_globalorbit}. Before the collision, gravity dominates the stream momentum and energy source. In the local simulations reported in this paper, the stream accurately follows the ballistic trajectory before collision, showing that the gas indeed is steered by gravity. 

In addition to the reported simulations, we also show the gas trajectory before the stream-stream collision in a test global calculation. The test simulation is performed with the same equations and scaling as listed in Section~\ref{sec:method_set-up}. However, instead of injecting streams at $\phi$ direction boundaries, we inject gas at $r=30.25\rs$ with velocity $\textbf{v}_{\rm inj}=(-0.158c,~0.0,~7.19\times10^{-2}c)$. The stream density is also calibrated to $f_{\rm Edd}=1.0$ , with initial temperature $T_{\rm inj}=10^{5}$K. The test calculation has spherical polar coordinate domain of $(1.1\rs,140.0\rs)\times(0,\pi)\times(0,2\pi)$. The root level is also resolved with $[64\times32\times64]$ cells. Three levels of static mesh refinement is applied to the region of $(1.2\rs,30.0\rs)\times(0.48\pi,0.52\pi)\times(0,2\pi)$ in $r,~\theta,~\phi$ direction. With this set-up, we cannot resolve the stream scale height $H$ at the injection site, according to estimation in Section~\ref{sec:method_set-up}. At the stream injection site, the stream cross-section is about $\sim200$ times larger than in the local simulations. 

Figure~\ref{fig:appendix_globalorbit} shows the $\theta=\pi/2$-plane gas density snapshot at $t'=0.27$, the cyan solid line shows the same orbit we adopt in the reported local simulations. The gas stream follows the ballistic trajectory well before collision, and is able to reproduce the collision angle and radius when intersecting with itself. The slight deviation from the ballistic trajectory after the pericenter is due to the slight angular momentum dissipation due to radiation force in the pericenter. We will elaborate the set-up detail and pursue further studies with similar global set-up in a subsequent paper.

\section{Resolution Study}\label{appendix:resolution}
\begin{figure}
    \centering \includegraphics[width=\textwidth]{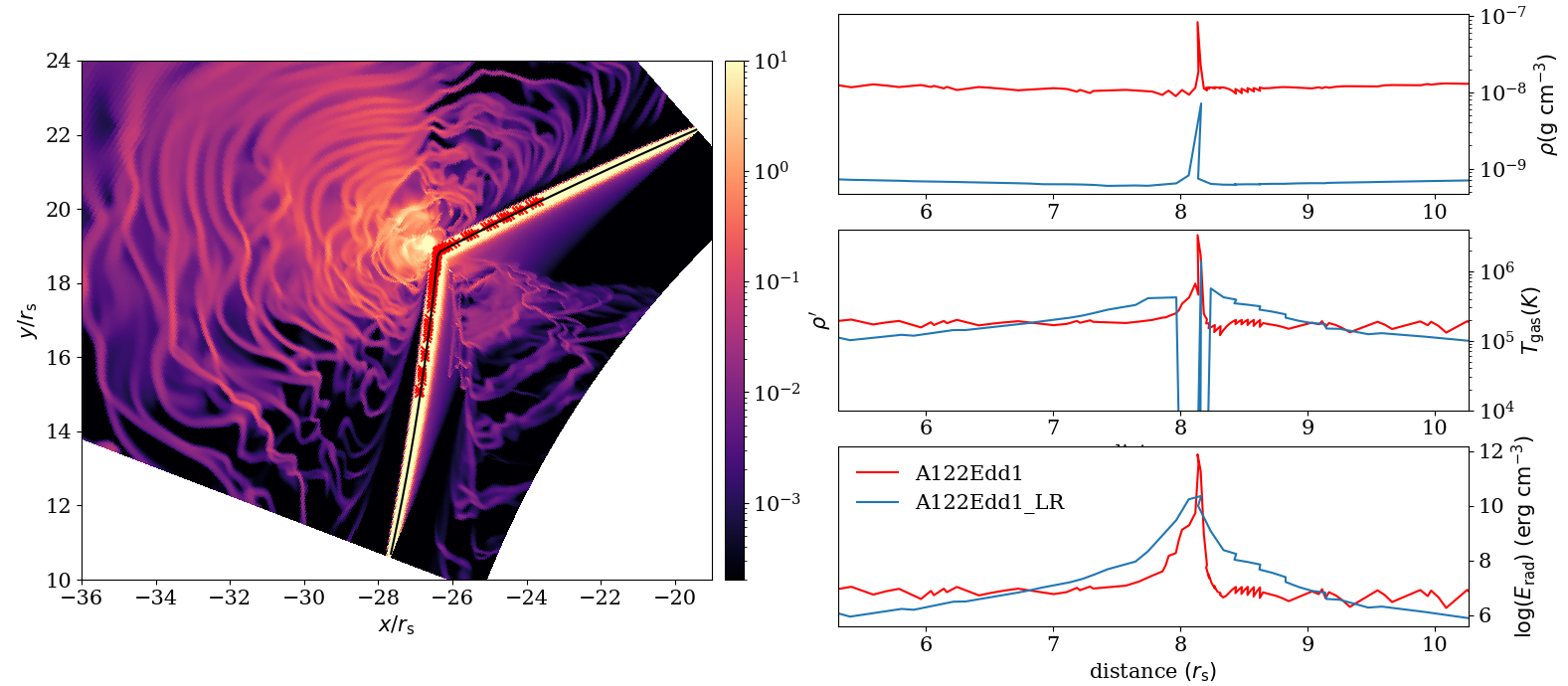}
    \caption{Left: gas density snapshot of A122Edd1 volume average from $\phi=1.56-1.59$ near $\theta=\pi/2$-plane, zoom-in view of Figure~\ref{fig:result_resolution_density} upper left panel. The black solid line is part of the ballistic trajectory, equivalent to the orbit shown in Figure~\ref{fig:method_domain}. Red cross data points are the sampled data points for the right column plots. Right: gas density (the first row), gas temperature (the second row) and radiation energy density (the third row) as a function of distance to the stream injection point at the inner $\phi$ boundary. The curves are plotted with the data sampled from the left panel along the stream before collision (indicated by the red cross points). The blue lines are for A122Edd1\_LR, the red lines are A122Edd1.}
    \label{fig:appendix_streamprofile}
\end{figure}

In order to understand better how resolution affects the radiation mediated shock at stream-stream collision site, Figure~\ref{fig:appendix_streamprofile} shows the gas and radiation profile along the two streams. In the left panel, we show the gas density snapshot for our fiducial resolution of A122Edd1, the black solid lines are part of the ballistic trajectory shown in Figure~\ref{fig:method_domain}.  It also shows that before the collision, the radiation energy and thermal energy are small compared to stream kinetic energy. 

Along the stream, we sample some data points (as shown in the red crosses) to study gas properties. From top to botoom, right panels in Figure~\ref{fig:appendix_streamprofile} show the gas density, gas temperature, and radiation energy density as a function of distance to the stream injection site at the left $\phi$ direction boundary.
The blue lines are for A122Edd\_LR and red lines are for A122Edd. The gas density is lower in the low resolution simulation due to the larger stream cross-section. The peak radiation energy density in A122Edd\_LR at the shock, however, is about two orders of magnitude lower than in A122Edd, suggesting that less kinetic energy is converted to radiation energy at the shock. The lower efficiency is likely due to both the under-resolved shock and the lower local optical depth in A122Edd\_LR.

We perform the simulations with the resolution specified in Section~\ref{sec:method_set-up}, where the streams before collision are resolved with five levels of SMR and the shock region is resolved with six levels of SMR, and  the radiation field is resolved with 48 angles. In the following resolution study, we present two more simulations: A122Edd\_R45 and A122Edd\_R45\_N80. Both simulations are with lower spatial resolution where the streams before collision are resolved with four levels of SMR and the shock region is resolved with five levels of SMR. However, we set $n_{\mu}=4$ angles per octant in A122Edd\_R45\_N80, so there are $n_{\rm oct}n_{\mu}(n_{\mu}+1)/2=80$ angles in total. 

\begin{figure}
    \centering
    \includegraphics[width=\textwidth]{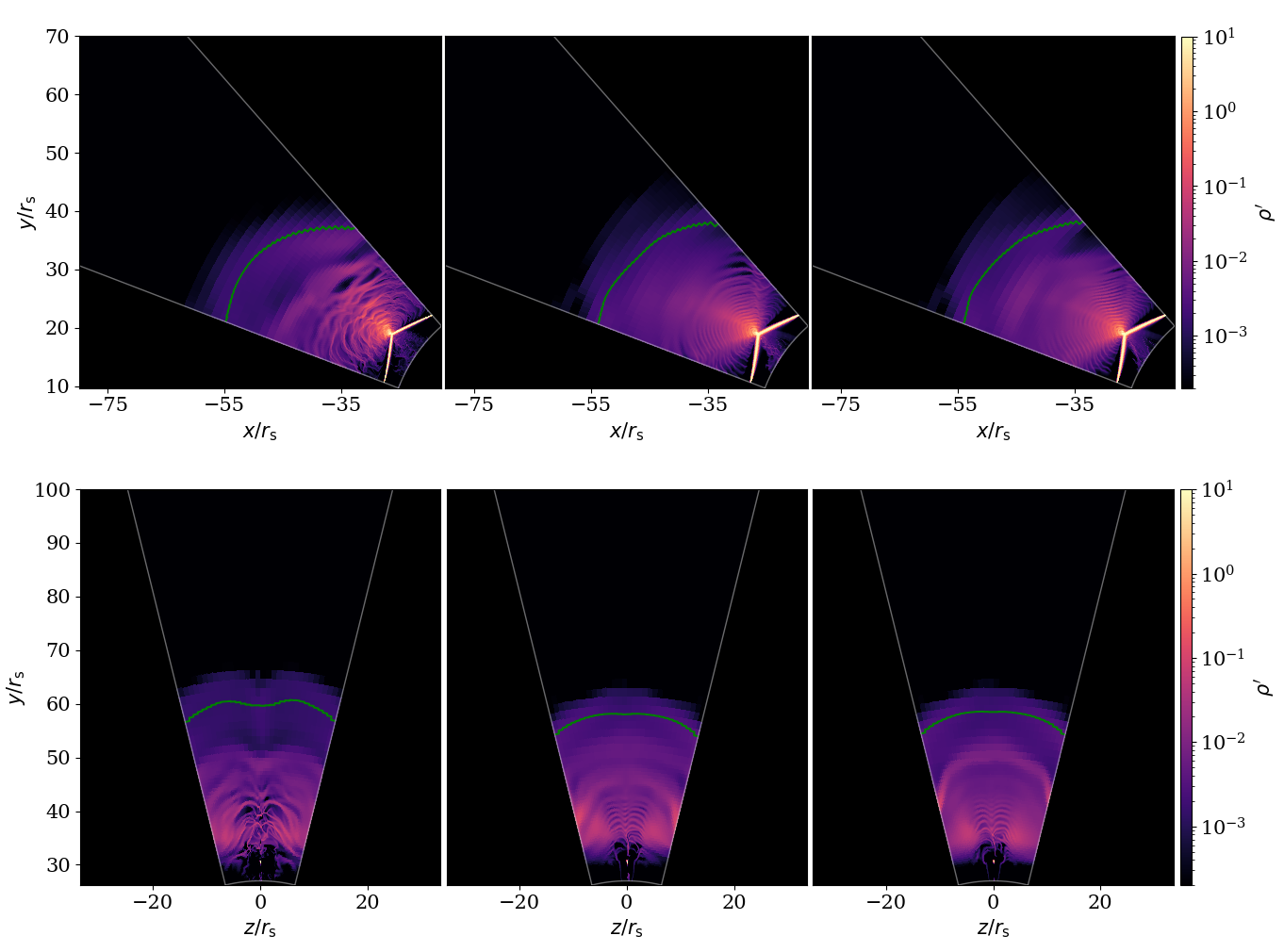}
    \caption{Similar to Figure~\ref{fig:result_resolution_density}, gas density snapshots at $t=0.3$ for A122Edd1 (the first column) A122Edd1\_R45 (the second column) and A122Edd1\_R45\_N80 (the third column). A122Edd1 is the fiducial simulation, A122Edd1\_R45 has one level lower of SMR compared to the fiducial simulation, A122Edd1\_R45\_N80 has one level lower of SMR, but higher angular resolution for radiation transfer. In each column, the upper plot shows volume average from $\phi=1.56-1.59$ near $\theta=\pi/2$-plane, the lower plot shows volume average from $\theta=2.52-2.55$ near mid-$\phi$-plane. The green solid line shows the same average for the photosphere radius $R_{\tau=1}$ as defined in Equation~\ref{eq:tau1}.}
    \label{fig:appendix_rf45_dens}
\end{figure}

Figure~\ref{fig:appendix_rf45_dens} compares density snapshots of these three simulations on different planes. The overall gas density distribution are similar. The lower spatial resolution (compare the second column to the first column) simulations produce less structure after the collision, but did not significantly change the photospheric radius. Increasing the number of discretization angles (compare the second column and the third column) in the radiation field has a only weak effect on the morphology of downstream gas.

\bibliography{ref}

\begin{thebibliography}{}
\expandafter\ifx\csname natexlab\endcsname\relax\def\natexlab#1{#1}\fi
\providecommand{\url}[1]{\href{#1}{#1}}
\providecommand{\dodoi}[1]{doi:~\href{http://doi.org/#1}{\nolinkurl{#1}}}
\providecommand{\doeprint}[1]{\href{http://ascl.net/#1}{\nolinkurl{http://ascl.net/#1}}}
\providecommand{\doarXiv}[1]{\href{https://arxiv.org/abs/#1}{\nolinkurl{https://arxiv.org/abs/#1}}}

\bibitem[{Andalman {et~al.}(2022)Andalman, Liska, Tchekhovskoy, Coughlin, \&
  Stone}]{andalman2022tidal}
Andalman, Z., Liska, M., Tchekhovskoy, A., Coughlin, E., \& Stone, N. 2022,
  AAS/High Energy Astrophysics Division, 54, 206

\bibitem[{Batra {et~al.}(2023)Batra, Lu, Bonnerot, \&
  Phinney}]{batra2023general}
Batra, G., Lu, W., Bonnerot, C., \& Phinney, E.~S. 2023, Monthly Notices of the
  Royal Astronomical Society, 520, 5192

\bibitem[{Ben-Ami {et~al.}(2022)Ben-Ami, Shvartzvald, Waxman, Netzer, Yaniv,
  Algranatti, Gal-Yam, Lapid, Ofek, Topaz, {et~al.}}]{ben2022scientific}
Ben-Ami, S., Shvartzvald, Y., Waxman, E., {et~al.} 2022, in Space Telescopes
  and Instrumentation 2022: Ultraviolet to Gamma Ray, Vol. 12181, SPIE, 13--23

\bibitem[{Bonnerot \& Lu(2022)}]{bonnerot2022nozzle}
Bonnerot, C., \& Lu, W. 2022, Monthly Notices of the Royal Astronomical
  Society, 511, 2147

\bibitem[{Bonnerot {et~al.}(2021)Bonnerot, Lu, \& Hopkins}]{bonnerot2021first}
Bonnerot, C., Lu, W., \& Hopkins, P.~F. 2021, Monthly Notices of the Royal
  Astronomical Society, 504, 4885

\bibitem[{Bonnerot {et~al.}(2022)Bonnerot, Pessah, \&
  Lu}]{bonnerot2022pericenter}
Bonnerot, C., Pessah, M.~E., \& Lu, W. 2022, The Astrophysical Journal Letters,
  931, L6

\bibitem[{Bonnerot {et~al.}(2017)Bonnerot, Price, Lodato, \&
  Rossi}]{bonnerot2017magnetic}
Bonnerot, C., Price, D.~J., Lodato, G., \& Rossi, E.~M. 2017, Monthly Notices
  of the Royal Astronomical Society, 469, 4879

\bibitem[{Bonnerot \& Stone(2021)}]{bonnerot2021formation}
Bonnerot, C., \& Stone, N. 2021, Space Science Reviews, 217, 1

\bibitem[{Bricman \& Gomboc(2020)}]{bricman2020prospects}
Bricman, K., \& Gomboc, A. 2020, The Astrophysical Journal, 890, 73

\bibitem[{{Chan} {et~al.}(2021){Chan}, {Piran}, \& {Krolik}}]{chanetal2021}
{Chan}, C.-H., {Piran}, T., \& {Krolik}, J.~H. 2021, \apj, 914, 107,
  \dodoi{10.3847/1538-4357/abf0a7}

\bibitem[{Coughlin(2023)}]{coughlin2023dynamics}
Coughlin, E.~R. 2023, Monthly Notices of the Royal Astronomical Society,
  stad1347

\bibitem[{Coughlin \& Begelman(2014)}]{coughlin2014hyperaccretion}
Coughlin, E.~R., \& Begelman, M.~C. 2014, The Astrophysical Journal, 781, 82

\bibitem[{Coughlin {et~al.}(2016)Coughlin, Nixon, Begelman, \&
  Armitage}]{coughlin2016structure}
Coughlin, E.~R., Nixon, C., Begelman, M.~C., \& Armitage, P.~J. 2016, Monthly
  Notices of the Royal Astronomical Society, 459, 3089

\bibitem[{Curd(2021)}]{curd2021global}
Curd, B. 2021, Monthly Notices of the Royal Astronomical Society, 507, 3207

\bibitem[{Dai {et~al.}(2015)Dai, McKinney, \& Miller}]{dai2015soft}
Dai, L., McKinney, J.~C., \& Miller, M.~C. 2015, The Astrophysical Journal
  Letters, 812, L39

\bibitem[{Dai {et~al.}(2018)Dai, McKinney, Roth, Ramirez-Ruiz, \&
  Miller}]{dai2018unified}
Dai, L., McKinney, J.~C., Roth, N., Ramirez-Ruiz, E., \& Miller, M.~C. 2018,
  The Astrophysical Journal Letters, 859, L20

\bibitem[{Davis {et~al.}(2012)Davis, Stone, \& Jiang}]{davis2012radiation}
Davis, S.~W., Stone, J.~M., \& Jiang, Y.-F. 2012, The Astrophysical Journal
  Supplement Series, 199, 9

\bibitem[{Dittmann(2022)}]{dittmann2022analytical}
Dittmann, A.~J. 2022, Monthly Notices of the Royal Astronomical Society, 511,
  3408

\bibitem[{Evans \& Kochanek(1989)}]{evans1989tidal}
Evans, C.~R., \& Kochanek, C.~S. 1989, The Astrophysical Journal, 346, L13

\bibitem[{Eyles-Ferris {et~al.}(2022)Eyles-Ferris, Starling, O’Brien, Nixon,
  \& Coughlin}]{eyles2022simulated}
Eyles-Ferris, R., Starling, R., O’Brien, P., Nixon, C., \& Coughlin, E.~R.
  2022, Monthly Notices of the Royal Astronomical Society, 517, 6013

\bibitem[{Gafton \& Rosswog(2019)}]{gafton2019tidal}
Gafton, E., \& Rosswog, S. 2019, Monthly Notices of the Royal Astronomical
  Society, 487, 4790

\bibitem[{Gehrels \& Cannizzo(2015)}]{gehrels2015journal}
Gehrels, N., \& Cannizzo, J. 2015

\bibitem[{Gezari(2021)}]{gezari2021tidal}
Gezari, S. 2021, Annual Review of Astronomy and Astrophysics, 59, 21

\bibitem[{Goldtooth {et~al.}(2023)Goldtooth, Zabludoff, Wen, Jonker, Stone, \&
  Cao}]{goldtooth2023census}
Goldtooth, A., Zabludoff, A.~I., Wen, S., {et~al.} 2023, Publications of the
  Astronomical Society of the Pacific, 135, 034101

\bibitem[{Greiner {et~al.}(2000)Greiner, Schwarz, Zharikov, \&
  Orio}]{greiner2000rx}
Greiner, J., Schwarz, R., Zharikov, S., \& Orio, M. 2000, arXiv preprint
  astro-ph/0009430

\bibitem[{Grupe {et~al.}(1999)Grupe, Thomas, \& Leighly}]{grupe1999rx}
Grupe, D., Thomas, H.-C., \& Leighly, K. 1999, arXiv preprint astro-ph/9909101

\bibitem[{Guillochon {et~al.}(2014)Guillochon, Manukian, \&
  Ramirez-Ruiz}]{guillochon2014ps1}
Guillochon, J., Manukian, H., \& Ramirez-Ruiz, E. 2014, The Astrophysical
  Journal, 783, 23

\bibitem[{Guillochon \& McCourt(2017)}]{guillochon2017simulations}
Guillochon, J., \& McCourt, M. 2017, The Astrophysical Journal Letters, 834,
  L19

\bibitem[{Guillochon \& Ramirez-Ruiz(2015)}]{guillochon2015dark}
Guillochon, J., \& Ramirez-Ruiz, E. 2015, The Astrophysical Journal, 809, 166

\bibitem[{Hambleton {et~al.}(2022)Hambleton, Bianco, Street, Bell, Buckley,
  Graham, Hernitschek, Lund, Mason, Pepper, {et~al.}}]{hambleton2022rubin}
Hambleton, K.~M., Bianco, F.~B., Street, R., {et~al.} 2022, arXiv preprint
  arXiv:2208.04499

\bibitem[{Hayasaki {et~al.}(2016)Hayasaki, Stone, \&
  Loeb}]{hayasaki2016circularization}
Hayasaki, K., Stone, N., \& Loeb, A. 2016, Monthly Notices of the Royal
  Astronomical Society, 461, 3760

\bibitem[{Hung {et~al.}(2017)Hung, Gezari, Blagorodnova, Roth, Cenko, Kulkarni,
  Horesh, Arcavi, McCully, Yan, {et~al.}}]{hung2017revisiting}
Hung, T., Gezari, S., Blagorodnova, N., {et~al.} 2017, The Astrophysical
  Journal, 842, 29

\bibitem[{Iglesias \& Rogers(1996)}]{iglesias1996updated}
Iglesias, C.~A., \& Rogers, F.~J. 1996, The astrophysical journal, 464, 943

\bibitem[{Jankovič {et~al.}(2023)Jankovič, Bonnerot, \&
  Gomboc}]{jankovic2023spininduced}
Jankovič, T., Bonnerot, C., \& Gomboc, A. 2023, Spin-induced offset stream
  self-crossing shocks in tidal disruption events.
\newblock \doarXiv{2303.16230}

\bibitem[{Jiang(2021)}]{jiang2021implicit}
Jiang, Y.-F. 2021, The Astrophysical Journal Supplement Series, 253, 49

\bibitem[{Jiang {et~al.}(2016)Jiang, Guillochon, \& Loeb}]{jiang2016prompt}
Jiang, Y.-F., Guillochon, J., \& Loeb, A. 2016, The Astrophysical Journal, 830,
  125

\bibitem[{Jiang {et~al.}(2014)Jiang, Stone, \& Davis}]{jiang2014global}
Jiang, Y.-F., Stone, J.~M., \& Davis, S.~W. 2014, The Astrophysical Journal,
  796, 106

\bibitem[{Kara {et~al.}(2018)Kara, Dai, Reynolds, \&
  Kallman}]{kara2018ultrafast}
Kara, E., Dai, L., Reynolds, C., \& Kallman, T. 2018, Monthly Notices of the
  Royal Astronomical Society, 474, 3593

\bibitem[{Kesden(2012)}]{kesden2012black}
Kesden, M. 2012, Physical Review D, 86, 064026

\bibitem[{Kim {et~al.}(1999)Kim, Park, \& Lee}]{kim1999stream}
Kim, S.~S., Park, M.-G., \& Lee, H.~M. 1999, The Astrophysical Journal, 519,
  647

\bibitem[{Kochanek(2016)}]{kochanek2016abundance}
Kochanek, C. 2016, Monthly Notices of the Royal Astronomical Society, 458, 127

\bibitem[{Kochanek {et~al.}(2017)Kochanek, Shappee, Stanek, Holoien, Thompson,
  Prieto, Dong, Shields, Will, Britt, {et~al.}}]{kochanek2017all}
Kochanek, C., Shappee, B., Stanek, K., {et~al.} 2017, Publications of the
  Astronomical Society of the Pacific, 129, 104502

\bibitem[{Komossa(2015)}]{komossa2015tidal}
Komossa, S. 2015, Journal of High Energy Astrophysics, 7, 148

\bibitem[{Law-Smith {et~al.}(2020)Law-Smith, Coulter, Guillochon, Mockler, \&
  Ramirez-Ruiz}]{law2020stellar}
Law-Smith, J.~A., Coulter, D.~A., Guillochon, J., Mockler, B., \& Ramirez-Ruiz,
  E. 2020, The Astrophysical Journal, 905, 141

\bibitem[{Lin {et~al.}(2011)Lin, Carrasco, Grupe, Webb, Barret, \&
  Farrell}]{lin2011discovery}
Lin, D., Carrasco, E.~R., Grupe, D., {et~al.} 2011, The Astrophysical Journal,
  738, 52

\bibitem[{Lin {et~al.}(2018)Lin, Strader, Carrasco, Godet, Grupe, Webb, Barret,
  \& Irwin}]{lin2018multiwavelength}
Lin, D., Strader, J., Carrasco, E.~R., {et~al.} 2018, Monthly Notices of the
  Royal Astronomical Society, 474, 3000

\bibitem[{Liptai {et~al.}(2019)Liptai, Price, Mandel, \&
  Lodato}]{liptai2019disc}
Liptai, D., Price, D.~J., Mandel, I., \& Lodato, G. 2019, arXiv preprint
  arXiv:1910.10154

\bibitem[{Lodato {et~al.}(2009)Lodato, King, \& Pringle}]{lodato2009stellar}
Lodato, G., King, A., \& Pringle, J. 2009, Monthly Notices of the Royal
  Astronomical Society, 392, 332

\bibitem[{Lodato \& Rossi(2011)}]{lodato2011multiband}
Lodato, G., \& Rossi, E.~M. 2011, Monthly Notices of the Royal Astronomical
  Society, 410, 359

\bibitem[{Loeb \& Ulmer(1997)}]{loeb1997optical}
Loeb, A., \& Ulmer, A. 1997, The Astrophysical Journal, 489, 573

\bibitem[{Lu \& Bonnerot(2020)}]{lu2020self}
Lu, W., \& Bonnerot, C. 2020, Monthly Notices of the Royal Astronomical
  Society, 492, 686

\bibitem[{MacLeod {et~al.}(2012)MacLeod, Guillochon, \&
  Ramirez-Ruiz}]{macleod2012tidal}
MacLeod, M., Guillochon, J., \& Ramirez-Ruiz, E. 2012, The Astrophysical
  Journal, 757, 134

\bibitem[{Metzger(2022)}]{metzger2022cooling}
Metzger, B.~D. 2022, The Astrophysical Journal Letters, 937, L12

\bibitem[{Metzger \& Stone(2016)}]{metzger2016bright}
Metzger, B.~D., \& Stone, N.~C. 2016, Monthly Notices of the Royal Astronomical
  Society, 461, 948

\bibitem[{Miller(2015)}]{miller2015disk}
Miller, M.~C. 2015, The Astrophysical Journal, 805, 83

\bibitem[{Mockler {et~al.}(2019)Mockler, Guillochon, \&
  Ramirez-Ruiz}]{mockler2019weighing}
Mockler, B., Guillochon, J., \& Ramirez-Ruiz, E. 2019, The Astrophysical
  Journal, 872, 151

\bibitem[{Phinney(1989)}]{phinney1989manifestations}
Phinney, E. 1989, in Symposium-International Astronomical Union, Vol. 136,
  Cambridge University Press, 543--553

\bibitem[{Piran {et~al.}(2015)Piran, Svirski, Krolik, Cheng, \&
  Shiokawa}]{piran2015disk}
Piran, T., Svirski, G., Krolik, J., Cheng, R.~M., \& Shiokawa, H. 2015, The
  Astrophysical Journal, 806, 164

\bibitem[{Ramirez-Ruiz \& Rosswog(2009)}]{ramirez2009star}
Ramirez-Ruiz, E., \& Rosswog, S. 2009, The Astrophysical Journal, 697, L77

\bibitem[{Rees(1988)}]{rees1988tidal}
Rees, M.~J. 1988, Nature, 333, 523

\bibitem[{Rosswog {et~al.}(2009)Rosswog, Ramirez-Ruiz, \&
  Hix}]{rosswog2009tidal}
Rosswog, S., Ramirez-Ruiz, E., \& Hix, W.~R. 2009, The Astrophysical Journal,
  695, 404

\bibitem[{Roth {et~al.}(2016)Roth, Kasen, Guillochon, \&
  Ramirez-Ruiz}]{roth2016x}
Roth, N., Kasen, D., Guillochon, J., \& Ramirez-Ruiz, E. 2016, The
  Astrophysical Journal, 827, 3

\bibitem[{{Rybicki} \& {Lightman}(1986)}]{1986rpa..book.....R}
{Rybicki}, G.~B., \& {Lightman}, A.~P. 1986, {Radiative Processes in
  Astrophysics}

\bibitem[{Ryu {et~al.}(2020)Ryu, Krolik, \& Piran}]{ryu2020measuring}
Ryu, T., Krolik, J., \& Piran, T. 2020, The Astrophysical Journal, 904, 73

\bibitem[{Sadowski {et~al.}(2016)Sadowski, Tejeda, Gafton, Rosswog, \&
  Abarca}]{sadowski2016magnetohydrodynamical}
Sadowski, A., Tejeda, E., Gafton, E., Rosswog, S., \& Abarca, D. 2016, Monthly
  Notices of the Royal Astronomical Society, 458, 4250

\bibitem[{Saxton {et~al.}(2012)Saxton, Read, Esquej, Komossa, Dougherty,
  Rodriguez-Pascual, \& Barrado}]{saxton2012tidal}
Saxton, R., Read, A., Esquej, P., {et~al.} 2012, Astronomy \& Astrophysics,
  541, A106

\bibitem[{Shiokawa {et~al.}(2015)Shiokawa, Krolik, Cheng, Piran, \&
  Noble}]{shiokawa2015general}
Shiokawa, H., Krolik, J.~H., Cheng, R.~M., Piran, T., \& Noble, S.~C. 2015, The
  Astrophysical Journal, 804, 85

\bibitem[{Spaulding \& Chang(2021)}]{spaulding2021effect}
Spaulding, A., \& Chang, P. 2021, Monthly Notices of the Royal Astronomical
  Society, 501, 1748

\bibitem[{Steinberg \& Stone(2022)}]{steinberg2022origins}
Steinberg, E., \& Stone, N.~C. 2022, arXiv preprint arXiv:2206.10641

\bibitem[{Stone \& Metzger(2016)}]{stone2016rates}
Stone, N.~C., \& Metzger, B.~D. 2016, Monthly Notices of the Royal Astronomical
  Society, 455, 859

\bibitem[{Strubbe \& Quataert(2009)}]{strubbe2009optical}
Strubbe, L.~E., \& Quataert, E. 2009, Monthly Notices of the Royal Astronomical
  Society, 400, 2070

\bibitem[{Tejeda \& Rosswog(2013)}]{tejeda2013accurate}
Tejeda, E., \& Rosswog, S. 2013, Monthly Notices of the Royal Astronomical
  Society, 433, 1930

\bibitem[{Thomsen {et~al.}(2022)Thomsen, Kwan, Dai, Wu, Roth, \&
  Ramirez-Ruiz}]{thomsen2022dynamical}
Thomsen, L.~L., Kwan, T.~M., Dai, L., {et~al.} 2022, The Astrophysical Journal
  Letters, 937, L28

\bibitem[{Ulmer(1999)}]{ulmer1999flares}
Ulmer, A. 1999, The Astrophysical Journal, 514, 180

\bibitem[{Van~Velzen {et~al.}(2011)Van~Velzen, Farrar, Gezari, Morrell,
  Zaritsky, {\"O}stman, Smith, Gelfand, \& Drake}]{van2011optical}
Van~Velzen, S., Farrar, G.~R., Gezari, S., {et~al.} 2011, The Astrophysical
  Journal, 741, 73

\bibitem[{Van~Velzen {et~al.}(2021)Van~Velzen, Gezari, Hammerstein, Roth,
  Frederick, Ward, Hung, Cenko, Stein, Perley, {et~al.}}]{van2021seventeen}
Van~Velzen, S., Gezari, S., Hammerstein, E., {et~al.} 2021, The Astrophysical
  Journal, 908, 4

\bibitem[{Wevers {et~al.}(2017)Wevers, van Velzen, Jonker, Stone, Hung, Onori,
  Gezari, \& Blagorodnova}]{wevers2017black}
Wevers, T., van Velzen, S., Jonker, P.~G., {et~al.} 2017, Monthly Notices of
  the Royal Astronomical Society, 471, 1694

\bibitem[{Wevers {et~al.}(2019)Wevers, Pasham, van Velzen, Leloudas, Schulze,
  Miller-Jones, Jonker, Gromadzki, Kankare, Hodgkin,
  {et~al.}}]{wevers2019evidence}
Wevers, T., Pasham, D., van Velzen, S., {et~al.} 2019, Monthly Notices of the
  Royal Astronomical Society, 488, 4816

\end{thebibliography}

\end{CJK*}
\end{document}